\documentclass[twocolumn,prd,aps,superscriptaddress,preprintnumbers,tightenlines,showpacs,nofootinbib,eqsecnum,amsfonts,amsmath]{revtex4}
\pdfoutput=1

\usepackage{graphicx,amssymb}
\usepackage{grffile}
\expandafter\let\csname equation*\endcsname\relax
\expandafter\let\csname endequation*\endcsname\relax
\usepackage{amsmath}
\usepackage{url}

\usepackage{float} 

\usepackage[utf8]{inputenc}

\usepackage{hyperref} 
\usepackage[varg]{txfonts}
\usepackage[T1]{fontenc}

\usepackage{algorithm}
\usepackage{algpseudocode}
\usepackage{longtable}
\usepackage{booktabs}
\usepackage{siunitx}

\usepackage{empheq,xcolor}
\definecolor{myblue}{rgb}{.8, .8, 1}

\DeclareMathAlphabet{\mathcalstd}{OMS}{cmsy}{m}{n}
\DeclareMathAlphabet{\mathpzc}{OT1}{pzc}{m}{it}

\newcommand{\Cardiff}{School of Physics and Astronomy, Cardiff University, Queens Building, CF24 3AA, Cardiff, United Kingdom}
\newcommand{\AEI}{Max Planck Institute for Gravitational Physics  (Albert Einstein Institute), Am M\"uhlenberg 1, D-14476 Potsdam-Golm, Germany}

\sloppypar

\begin{document}


\title{Frequency domain reduced order model of aligned-spin effective-one-body waveforms\\ 
with generic mass-ratios and spins}

\author{Michael P\"urrer}
\affiliation{\Cardiff}
\affiliation{\AEI}

\begin{abstract}
I provide a frequency domain reduced order model (ROM) for the aligned-spin effective-one-body (EOB) model ``SEOBNRv2'' for data analysis with second and third generation ground based gravitational wave (GW) detectors. SEOBNRv2 models the dominant mode of the GWs emitted by the coalescence of black hole (BH) binaries. The large physical parameter space (dimensionless spins $-1 \leq \chi_i \leq 0.99$ and symmetric mass-ratios $0.01 \leq \eta \leq 0.25$) requires sophisticated reduced order modeling techniques, including patching in the parameter space and in frequency. I find that the time window over which the inspiral-plunge and the merger-ringdown waveform in SEOBNRv2 are connected is discontinuous when the spin of the deformed Kerr BH $\chi=0.8$ or the symmetric mass-ratio $\eta \sim 0.083$. This discontinuity increases resolution requirements for the ROM. The ROM can be used for compact binary systems with total masses of $2 M_\odot$ or higher for the advanced LIGO (aLIGO) design sensitivity and a $10$ Hz lower cutoff frequency. The ROM has a worst mismatch against SEOBNRv2 of $\sim 1\%$, but in general mismatches are better than $\sim 0.1\%$. The ROM is crucial for key data analysis applications for compact binaries, such as GW searches and parameter estimation carried out within the LIGO Scientific Collaboration (LSC).
\textbf{LIGO-P1500175}
\end{abstract}


\maketitle


\section{Introduction} 
\label{sec:introduction}

The Advanced LIGO (aLIGO) gravitational wave (GW) detectors~\cite{Advanced-LIGO2015} have recently started their first observing run and Advanced Virgo~\cite{Advanced-VIRGO2015} is expected to start taking data in 2016. The first direct detection of gravitational waves is expected to be made before the end of this decade when these detectors should reach their design sensitivity~\cite{Aasi:2013wya}.
The Einstein Telescope is a planned third generation detector and will be an order of magnitude more sensitive that second generation detectors once it becomes operational after 2025~\cite{Punturo:2010zza}.

Coalescences of binary black hole (BH) systems are one of the most promising sources for ground based gravitational wave detectors. 
Possible GW signals will be buried in the noisy output of the interferometers. Matched filtering can be used to look for signals in GW searches by correlating the detector data against GW templates~\cite{Allen:2005fk}. Parameter estimation then attempts to extract more accurate information on astrophysical parameters of the binary.

A prerequisite for searches and especially parameter estimation is the availability of fast and accurate models of the GW waveform emitted from BH binaries so as not to miss signals or misrepresent their astrophysical parameters. Typically $10^6 - 10^7$ waveform model evaluations are needed for a PE simulation. 
Stochastic template banks algorithms~\cite{Harry:2009ea,Manca:2009xw} for GW searches typically produce banks up to $N \sim 5 \times 10^5$ templates~\cite{PhysRevD.89.084041,PhysRevD.89.024003,PhysRevD.89.024010} in the aligned-spin parameter space. In the construction of these banks many more waveforms need to be generated. The authors of~\cite{Harry:2009ea} suggest that $N^2$ evaluations are used to make sure the final bank is effectual. This number of evaluations is not feasible even for the fastest waveform models and the number used in practise is estimated to be $\sim N^{1.5} \approx 10^8$ waveform evaluations~\cite{IHcomment}.
While the construction of the bank is very costly, it can be used as long as the power spectral density (PSD) of the detector noise does not change too much~\cite{Usman:2015kfa}.

Inspiral-merger-ringdown models of GWs emitted from BH binaries have been constructed either in the effective-one-body (EOB) approach~\cite{Buonanno:1998gg,Buonanno:2000ef,Damour:2000we,Damour:2001tu,Damour:2008gu} or as phenomenological models~\cite{Husa:2015iqa,Khan:2015jqa,Ajith:2011,Santamaria:2010yb}. In this paper I focus on the ``SEOBNRv2''~\cite{Taracchini:2013rva} model that describes GWs from binaries with spins aligned with the orbital angular momentum. Compared to its predecessor ``SEOBNRv1''~\cite{Taracchini:2012ig} it extends the spin range to almost the entire Kerr range.

EOB waveforms are obtained from the integration of complicated systems of ordinary differential equations and are in general much too slow for direct data analysis applications. Reduced order modeling (ROM) can provide fast and accurate surrogates for such GW models which are crucial for GW searches and parameter estimation. I describe here a ROM that speeds up the generation of SEOBNRv2 waveforms by a factor of thousands. This allows parameter estimation studies to be carried out on the order of a day instead of a year.
ROM approaches are based either on the singular value decomposition (SVD) and different interpolation techniques~\cite{Purrer:2014fza,Smith:2012du,Cannon:2012gq,Cannon:2011rj} or on the greedy basis method and empirical interpolation~\cite{Field:2013cfa, Field:2011mf, Blackman:2014maa, Blackman:2015pia}.

In this paper I extend the reduced order modeling techniques presented in~\cite{Purrer:2014fza} to SEOBNRv2 waveforms and provide a ROM that covers the \emph{entire} parameter space domain on which SEOBNRv2 is defined and a frequency range sufficient for \emph{all} expected compact binary sources including binary BHs, neutron star - black hole, and binary neutron star systems, while neglecting equation of state effects.
A unified ROM covering the mass space for all compact binary coalescences is of great utility for data analysis, even though SEOBNRv2 models GWs emitted from BH binaries and does not include tidal effects or describe tidal disruption. In principle, such effects could be added post hoc as modifications of the amplitude and phasing of SEOBNRv2~\cite{Pannarale:2015jka,Lackey:2013axa}.
The ROM provides speedups of up to several thousands over the SEOBNRv2 model while being accurate to better than $1\%$ in the mismatch to keep ROM errors below the calibration accuracy of SEOBNRv2 against NR waveforms~\cite{Taracchini:2013rva}.

The approach to build frequency domain reduced order (or surrogate) models presented in~\cite{Purrer:2014fza} contains the following important steps:
EOB input waveforms are generated on a tensor product grid over the domain of intrinsic parameters (mass-ratio, aligned spins). The waveforms are tapered and Fourier transformed and split into their amplitude and phase parts. These functions can be accurately represented on sparse frequency grids with only O(100) nodes. The total mass does not need to be a parameter when working with geometric frequency $Mf$. I build separate reduced bases for the amplitude and phase with the \emph{singular value decomposition} (SVD).
To construct a waveform model the expansion coefficients of the amplitude and phase as a function of physical parameters are needed. I compute these coefficients from projecting the input amplitudes and phases onto their reduced bases and interpolate the coefficients by tensor product cubic splines. Finally I assemble the ROM from its constituent parts and compute the strain from the amplitude and phase interpolants.

In this paper I extend these techniques for the more challenging SEOBNRv2 case. In particular, the aligned spin range is extended to go close to extremal aligned spins, $\chi_i \in [-1, 0.99]$.
As will be discussed later this extended region turns out to be more challenging to model than expected due to peculiarities of the SEOBNRv2 model and it serves as a nice illustratation of how ROM methods can help find problems in waveform models.
Crucially, to obtain a faithful model over a large parameter space I employ \emph{domain decomposition techniques} both in the mass-ratio and spins and in frequency to increase resolution at high spins and for high frequencies. 
Domain decomposition methods and ``hp-greedy algorithms'' for adaptive sampling for large problems have been discussed in Refs.~\cite{Field:2013cfa,Fares20115532,doi:10.1080/13873954.2011.547670,Eftang-2stepRB,Eftang-parameter-multi-domain}.

The organization of this paper is as follows: In section~\ref{sec:the_seobnrv2_model} I summarize the SEOBNRv2 waveform model. In section~\ref{sec:reduced_order_modeling_techniques} I describe new reduced order modeling techniques introduced in this paper and discuss in section~\ref{sec:anatomy_of_the_seobnrv2_rom} how the SEOBNRv2 ROM is constructed. Results on speedup and accuracy of the ROM are discussed in section~\ref{sec:speedup_and_accuracy}. Conclusions are given in section~\ref{sec:conclusions}.


\section{The SEOBNRv2 model} 
\label{sec:the_seobnrv2_model}

The gravitational wave signal emitted from coalescing binary black holes can be characterized by three different stages. During the long inspiral the orbital velocities are small compared to the speed of light and this stage is well described by post-Newtonian theory~\cite{Blanchet:2013haa}. Driven by gravitational radiation reaction the companions approach each other more and more and eventually merge. In that regime, non-perturbative solutions of Einstein's equations are needed which can be computed by numerical relativity simulations~\cite{Centrella:2010mx}. The ringdown of the remnant Kerr black hole and the emission of quasinormal modes can be described by perturbation theory~\cite{Berti:2009kk}.

The effective-one-body formalism introduced by Buonanno and Damour~\cite{Buonanno:1998gg,Buonanno:2000ef,Damour:2000we,Damour:2001tu,Damour:2008gu,Buonanno:2014aza} combines post-Newtonian theory, resummation techniques and perturbation theory with the help of mapping the two body problem to an effective one body problem.
This approach assumes that the comparable-mass behavior is a smooth deformation of the test-particle limit.
Three main ingredients of EOB are the conservative two-body dynamics, radiation-reaction, and the gravitational waveform emitted during inspiral, merger and ringdown.

The full EOB waveform is obtained by joining the inspiral-plunge waveform and the merger-rindown waveform over a predetermined interval in time. Over the years the formalism has evolved to incorporate spin orbit and spin-spin interactions, and several high order parameters are calibrated from NR information~\cite{Damour:2009kr,Pan:2011gk}.

In the following I summarize the aligned-spin waveform model ``SEOBNRv2'' introduced by Taracchini et al.~\cite{Taracchini:2013rva} as an extension of an earlier ``SEOBNRv1''~\cite{Taracchini:2012ig} model that was limited to dimensionless spins smaller than $0.6$.
SEOBNRv2 is a model for the dominant $l=m=2$ GW mode emitted from BH binary systems with spins aligned with the orbital angular momentum. The dimensionless component spins $\chi_i = \vec S_i \cdot \hat L / m_i^2$ cover almost the full Kerr range $-1 \leq \chi_i \leq 0.99$. $\vec S_i$ are the spin vectors, $\hat L$ the orbital angular momentum unit vector and $m_i$ are the component masses of the BHs. The mass-ratio $q = m_1/m_2 \geq 1$ is allowed to range between equal mass and quite extreme systems at $q=100$. 
In terms of the symmetric mass-ratio $\eta := m_1 m_2 / (m_1 + m_2)^2$ the model covers approximately $0.01 \leq \eta \leq 0.25$.

\begin{figure}[htbp]
  \centering
    \includegraphics[width=.49\textwidth]{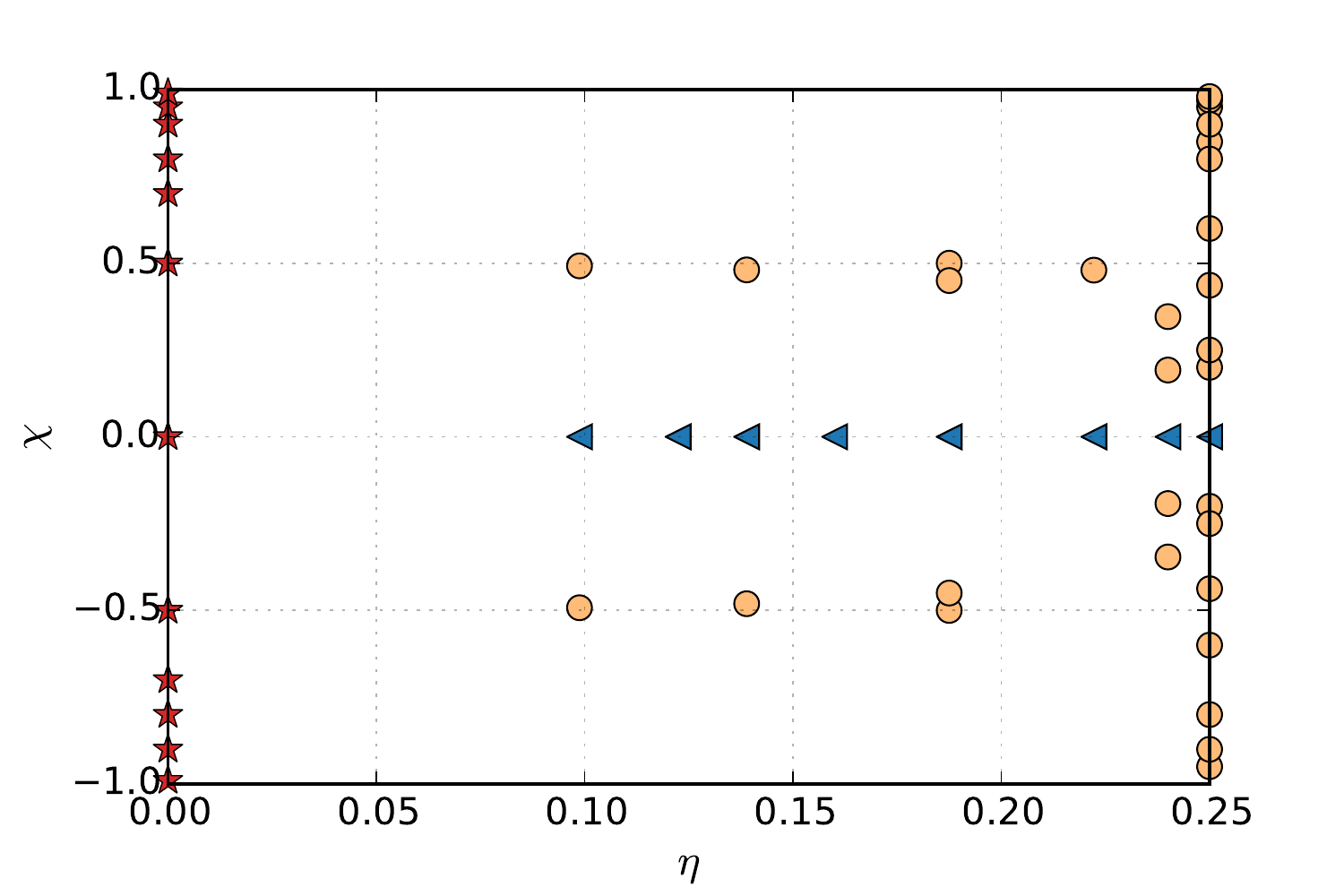}
  \caption{The 38 SXS nonspinning (blue) and spinning (orange) NR simulations used to calibrate SEOBNRv2 as a function of symmetric mass-ratio $\eta$ and the spin parameter $\chi$. In addition, the Teukolsky waveforms are shown in red.}
  \label{fig:plots_NRSimulationsv2}
\end{figure}

The model was calibrated against the $l=m=2$ mode of a catalog of 38 BH binary NR waveforms produced by the SXS Collaboration~\cite{Mroue:2013xna, Mroue:2012kv, PhysRevD.88.064014, Hemberger:2012jz}. In particular, 8 nonspinning NR waveforms with $q = 1, 1.5, 2, 3, 4, 5, 6, 8$, plus 19 spinning NR waveforms were used for calibration. In addition the model was tuned against Teukolsky-based merger-waveforms for $\eta = 0.001$ and several spin values for the central BH. These calibration points are shown in Fig.~\ref{fig:plots_NRSimulationsv2} as a function of the symmetric mass-ratio and a spin parameter
\begin{equation}
  \label{eq:chi_bg}
  \chi := \frac{\chi_\mathrm{Kerr}}{1 - 2\eta} = \frac{\chi_1 + \chi_2} {2} + \left(\frac{\chi_1 - \chi_2}{2}\right)
  \left(\frac{m_1 - m_2}{m_1 + m_2}\right) \left(\frac{1}{1-2\eta}\right),
\end{equation}
where
$\chi_\mathrm{Kerr} = (\vec S_1 + \vec S_2)\cdot \hat L / M^2$ is the dimensionless spin of the deformed Kerr geometry used in the EOB construction.

The model was found to be accurate to a mismatch of $1\%$ against the same set of SXS NR waveforms for total masses in $[20, 200] M_\odot$ and the aLIGO design power spectral density (PSD)~\cite{T0900288} with a low frequency cutoff of that depends on the length of each NR waveform (see Fig. 1 in Ref.~\cite{Taracchini:2013rva}). A study focused on NSBH systems~\cite{Kumar:2015tha} also found mismatches of $1\%$ or better between SEOBNRv2 and NR waveforms.

A recent study found evidence~\cite{Khan:2015jqa} that while SEOBNRv2 is in general extremely accurate it may not accurately describe the merger-ringdown regime for high aligned spins $\chi_i \gtrsim 0.7$ where it was not calibrated to NR simulations. This is corroborated by a further waveform comparison study~\cite{Prayush-Chu-SEOBNR-Phenom-comparison} with a catalog of new SXS NR waveforms~\cite{SXS-Chu-catalog}.

\section{Reduced order modeling techniques} 
\label{sec:reduced_order_modeling_techniques}

In the following I discuss extensions to the reduced order modeling approach given in~\cite{Purrer:2014fza}. In particular, I focus on techniques to achieve increased resolution in some parts of the model parameter space and on an efficient way of handling different resolution requirements over the frequency domain.
This allows building accurate models at an economical cost.

As discussed in~\cite{Purrer:2014fza}, I build separate bases for the frequency domain amplitude and phasing of the original waveform model using the singular value decomposition. To obtain a ROM a prediction of amplitude and phase coefficients is needed along with the reduced bases. This can be achieved by interpolating the coefficients over the parameter space. Tensor product splines are a robust tool to achieve this, if the density of waveforms in the multi-dimensional regular grid is sufficiently high to capture the variations in the waveform model.

The final product is a surrogate for the GW strain for the $l=m=2$ mode in the frequency domain~\cite{Purrer:2014fza}
\begin{multline}
	\tilde h(\lambda; M; f) := A_0(\lambda,M) \;
		I_f\left[ \mathcal{B}_\mathcal{A} \cdot I_\otimes[\mathcal{M}_\mathcal{A}](\lambda) \right] \\
		\exp\left\{ i \, I_f\left[ \mathcal{B}_\Phi \cdot I_\otimes[\mathcal{M}_\Phi](\lambda) \right] \right\},
\end{multline}
with the intrinsic model parameters $\lambda = \{ \eta, \chi_1, \chi_2 \}$, the reduced bases $\mathcal{B}$, projection coefficient tensors $\mathcal{M}$, tensor product $I_\otimes []$ and frequency spline interpolation operators $I_f []$. The two polarizations of the dominant $l=m=2$ mode of the GW strain are then given by Eq. (6.14) and (6.15) of Ref.~\cite{Purrer:2014fza}.

\subsection{Patching in the parameter space} 
\label{sub:patching_in_the_parameter_space}

\begin{figure}[htbp]
  \centering
    \includegraphics[width=.49\textwidth]{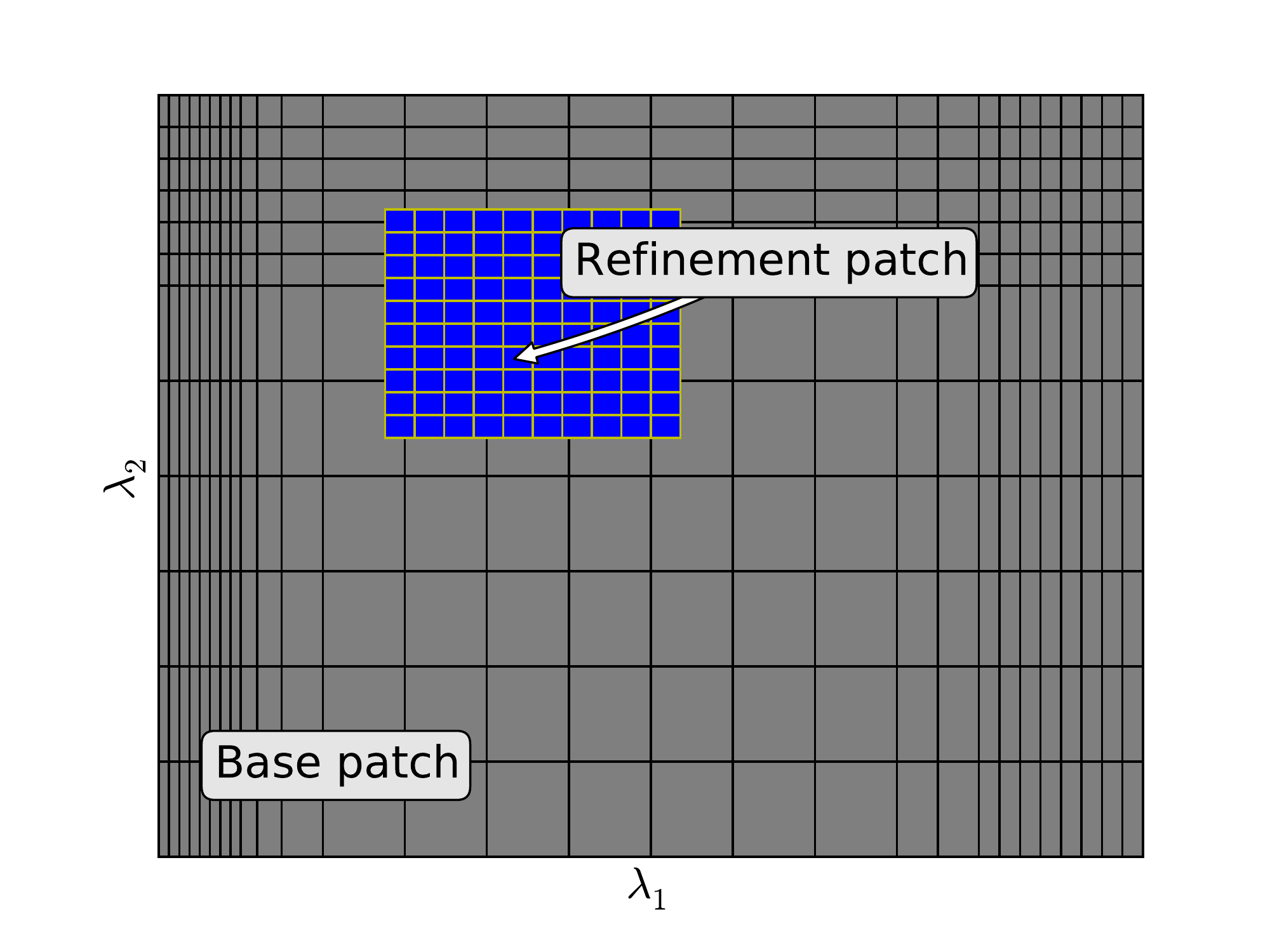}
  \caption{An example tensor product grid on a base patch covering the entire domain of interest in model parameters $\lambda_1, \lambda_2$ and a refinement patch.}
  \label{fig:plots_grid_patches}
\end{figure}

I will show in Sec.~\ref{sub:non_smoothness_in_seobnrv2_at_for_high_spin} that resolution requirements for SEOBNRv2 are not uniform over the parameter space. Tensor product interpolation does not allow for local refinement regions, as it is defined as the tensor product of one-dimensional sets of points. A simple improvement can be obtained by decomposing the domain into multiple overlapping patches. This idea is illustrated in Fig.~\ref{fig:plots_grid_patches}. A base patch, covering the whole domain of interest, is complemented by one (or more) smaller refinement patches for regions of higher waveform variability. These regions can be found quite naturally when checking the accuracy of a ROM covering the whole domain.

The combined ROM then switches between ROMs defined on subdomains based on the location where the ROM is to be evaluated. It is prudent to switch between these sub-ROMs not right at the boundary, but somewhat inside the subdomain in question to avoid possible boundary effects. The subdomains chosen for the SEOBNRv2 ROM are described in Sec.~\ref{sec:anatomy_of_the_seobnrv2_rom}.

\subsection{Patching in frequency} 
\label{sub:patching_in_frequency}

Resolution requirements also depend on frequency. Due to highly non-linear physics near merger and calibration against NR waveforms it is natural to assume that SEOBNR models will contain more structure at high frequencies, while the behavior at low frequency is expected to be very smooth based on the resummed post-Newtonian information used in the construction of the EOB model. This can indeed be seen by studying the performance of a ROM which covers a broad frequency range at different total masses with an appropriate detector power spectral density. In particular I find that SEOBNRv2 does not smoothly depend on mass-ratio and spins everywhere in the parameter space (see Sec.~\ref{sub:non_smoothness_in_seobnrv2_at_for_high_spin} for details).

Since the cost of EOB waveform generation rises steeply as the starting frequency is decreased, the above resolution requirements imply that it is efficient to split the frequency domain into two parts. One ROM can cover the low frequency part with reduced resolution over the parameter space, while another can explore the finer structure present at high frequencies with a high density of waveforms that can be generated in an affordable manner. For the present work these high frequency waveforms are generated at a total mass of $100 M_\odot$. In principle, the frequency domain could be partitioned into more than two subdomains, but since each subdomain introduces overhead costs due to interpolation over the parameter space at model evaluation, I restrict myself to two parts.

The input waveforms used to build a ROM using the method outlined in~\cite{Purrer:2014fza} must be defined on a fixed grid in geometric frequency. In particular, there is a fixed geometric starting frequency determined from the lowest total mass of interest, $2 M_\odot$ for BNS systems, and from the low frequency cutoff of the detector noise spectrum, which is close to $10$ Hz for aLIGO design. For configurations with high mass-ratio this would imply component masses way below $1 M_\odot$, a lower bound for the expected mass of a neutron star, and thus these systems are of no astrophysical interest for GW astronomy. There is no need to produce SEOBNRv2 waveforms for thsese configurations at such low frequency, but since the input waveforms need to fill the entire geometric frequency domain of interest, some data is required there. I generate the waveforms at a fixed mass of the smaller body of $1 M_\odot$ and then hybridize them with the TaylorF2 stationary phase approximant. The hybridization procedure is described in Appendix A of~\cite{Purrer:2014fza}.

At the high frequency end, the variation in ringdown frequency of SEOBNRv2 causes the ringdown for some configurations to start at a frequency where the ringdown has already stopped for others in the Fourier transformed data. To prevent problems with extrapolation I fit the ringdown in the frequency domain for all input waveforms, as described in Sec.~\ref{sub:treating_the_ringdown}.

To obtain an overall ROM the low and high frequency ROMs can be glued together in frequency as follows.
The phasing is glued over an overlapping frequency interval with the help of least squares fits to be $C^1$. This is discussed in Sec.~\ref{sub:gluing_the_roms_in_frequency}. For the amplitude I have found it sufficient to simply connect the two discrete amplitude functions at a chosen frequency in the frequency interval where they overlap.
On each frequency subdomain the discrete amplitude and phase functions are given by evaluating
$\mathcal{B} \cdot I_\otimes[\mathcal{M}](\lambda)$ for a specified point $\lambda = \{ \eta, \chi_1, \chi_2 \}$ in the parameter space.




\section{Anatomy of the SEOBNRv2 ROM} 
\label{sec:anatomy_of_the_seobnrv2_rom}

Here I discuss in detail the subdomains in frequency and model parameters $\eta, \chi_1, \chi_2$ from which the overall ROM is constructed. I use a single domain covering the entire parameter space of SEOBNRv2 at low frequency. The parameter space is split into two subregions at high frequency, with a transition at the spin of the larger BH $\chi_1 = 0.41$. Both subregions have some overlap around this $\chi_1$ value. The final ROM is assembled from these three sub models.

The gluing of ROMs in frequency is discussed in Sec.~\ref{sub:gluing_the_roms_in_frequency}, the treatment of the ringdown in Sec.~\ref{sub:treating_the_ringdown} and the ROMs on the low and high frequency subdomains are described in Sec.~\ref{sub:low_mass_rom} and Sec.~\ref{sub:high_mass_high_chi1_rom}.

\subsection{Gluing the ROMs in frequency} 
\label{sub:gluing_the_roms_in_frequency}

I choose a transition frequency of $Mf_m = 0.01$ where the low and high frequency ROMs will be connected so that waveforms generated at a total mass of $100 M_\odot$ are of sufficient length. The rationale for patching in frequency has been discussed in Sec.~\ref{sub:patching_in_frequency}.

The phasing of the low frequency ROM is first interpolated onto the grid of the high frequency ROM. A cubic polynomial is then fitted to the phasing of the low and high frequency ROM in a frequency interval $Mf \in [0.0074, 0.014]$ covered by 15 sparse frequency points below and above the gluing frequency. I then compute
\begin{align}
  \Delta\omega_m  &= \frac{d\phi_\text{high}}{df}\bigg\vert_{Mf_m} - \frac{d\phi_\text{low}}{df}\bigg\vert_{Mf_m}\\
  \Delta\phi_m    &= \phi_\text{high} (Mf_m) - \phi_\text{low} (Mf_m) - \Delta\omega_m \, Mf_m\\
\end{align}
and correct the phasing of the high frequency ROM by
\begin{equation}
  \phi_\text{high}(f) - \Delta\omega_m \, f - \Delta\phi_m.
\end{equation}
$\phi_\text{low}(f)$ is used unchanged up to $Mf_m$.


\subsection{Extending the ringdown} 
\label{sub:treating_the_ringdown}

\begin{figure*}[htbp]
  \centering
    \includegraphics[width=0.49\textwidth]{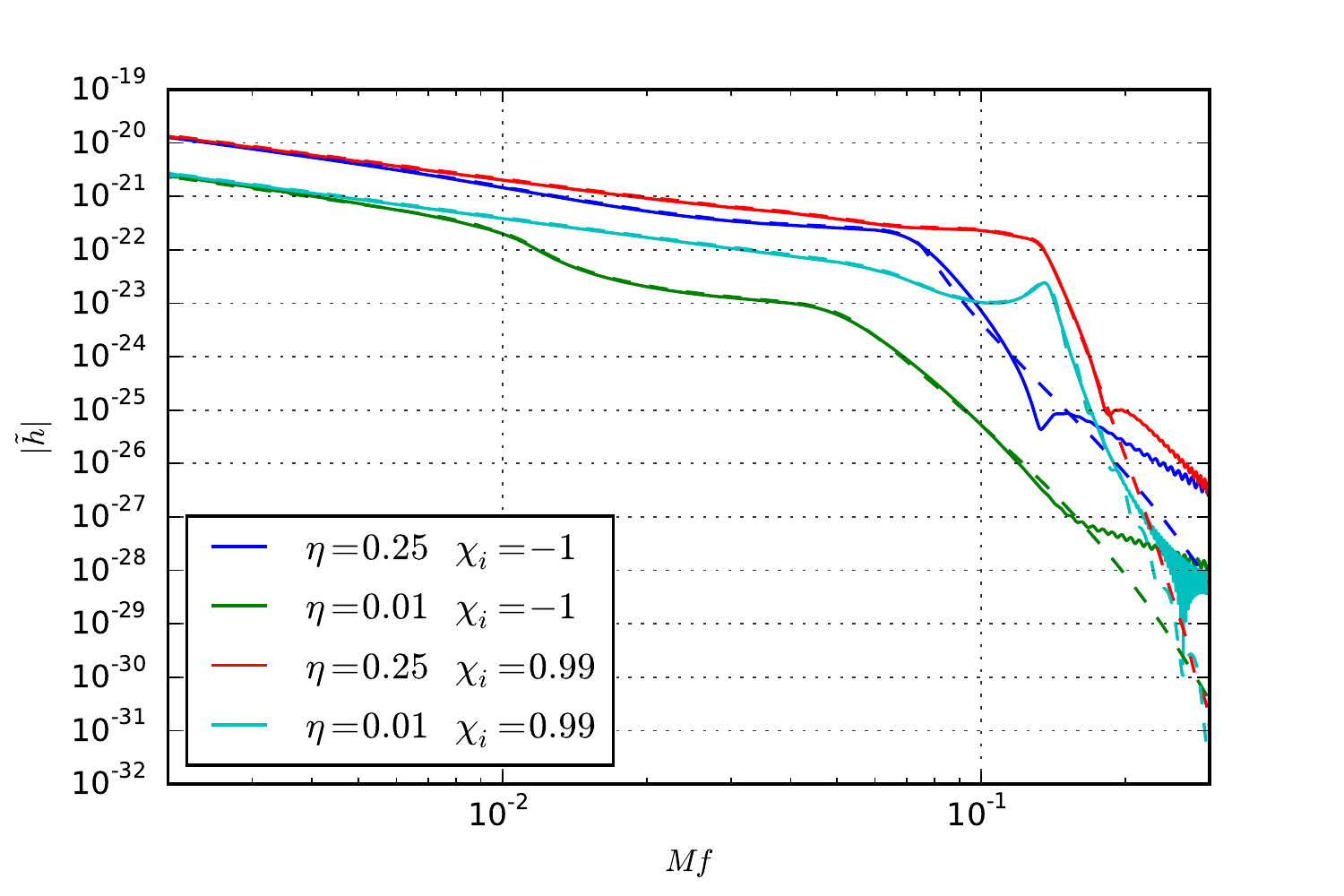}
      \includegraphics[width=.49\textwidth]{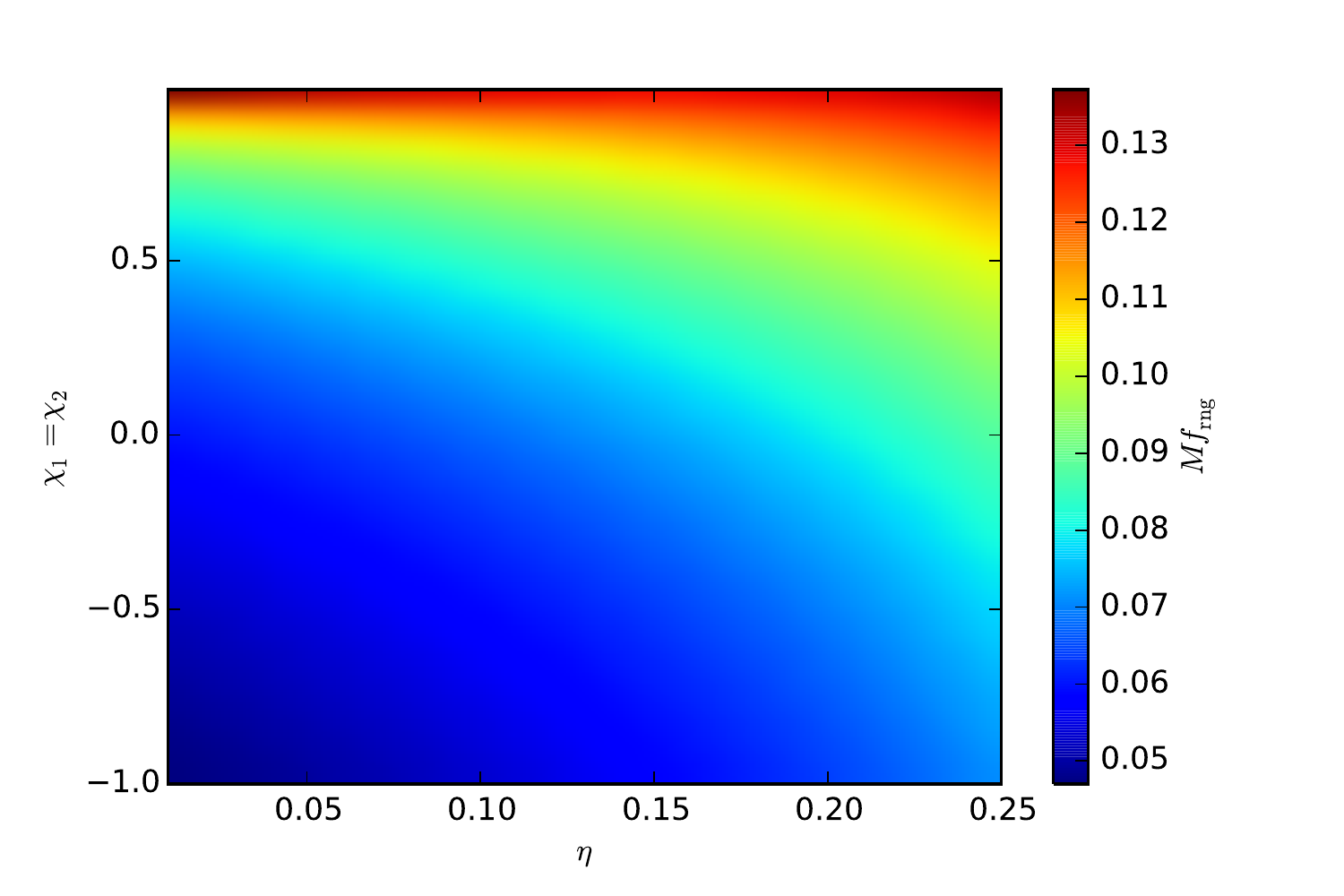}
  \caption{The Fourier domain amplitude for four corner cases in the parameter space (left panel). The FFT of SEOBNRv2 (solid) and the ROM (dashed) are shown. The SEOBNRv2 ringdown for extreme anti-aligned configurations $\chi_i = -1$ terminates earlier than the beginning of the ringdown for $chi_i=0.99$. 
  The ringdown frequency of SEOBNRv2 as a function of the symmetric mass-ratio $\eta$ and equal spin $\chi_1 = \chi_2$ (right panel).
  }
  \label{fig:amplitude-corner-cases}
\end{figure*}

The ROM construction in geometric frequency requires the waveform data to be given on a common frequency grid. As Fig.~\ref{fig:amplitude-corner-cases} shows, the ringdown for highly aligned waveforms starts at a frequency where the ringdown in the FFT data of highly anti-aligned waveforms has already terminated. One solution is to extend the ringdown data from the FFT of SEOBNRv2 by a fit and use this as \emph{input} data for building the ROM. In particular, I fit the ringdown amplitude of all input waveforms to a Lorentzian function~\cite{Khan:2015jqa}
\begin{equation}
  \mathcal{L}(Mf) = \gamma_1
  \frac{ (\gamma_3 Mf_\text{DM})^2 } {(Mf - Mf_\text{RD})^2 + (\gamma_3 Mf_\text{DM})^2}
  e^\frac{-\gamma_2 (Mf - Mf_\text{RD})}{\gamma_3 Mf_\text{DM}},
\end{equation}
The ringdown frequency $Mf_\text{RD}$ and damping frequency $Mf_\text{DM}$ are computed from the LAL code~\cite{LAL-web} for SEOBNRv2. $\gamma_1$ is set to the amplitude at the ringdown frequency $Mf_\text{RD}$ and I fit $\gamma_2$ and $\gamma_3$ in the frequency interval $Mf \in [Mf_\text{RD}, 0.18]$.
I use the fits to extend the ringdown amplitude from the ringdown frequency $Mf_\text{RD}$ until $Mf = 0.3$ so that the ringdown part is represented for all configurations in the domain of definition of SEOBNRv2. Linear extrapolation in the phase is sufficient due to the rapid decrease in the amplitude. Since the extrapolation of the ringdown is applied to all SEOBNRv2 input waveforms, the ROM automatically incorporates this information and can be evaluated up to $Mf = 0.3$.

\subsection{Low frequency ROM} 
\label{sub:low_mass_rom}

I build a low frequency ROM that covers the entire SEOBNRv2 parameter space $0.01 \leq \eta \leq 0.25$, $-1 \leq \chi_i \leq 0.99$. The waveforms are generated at a sampling rate of $32768$ Hz with a length sufficient for systems of a total mass of $2 M_\odot$ or larger. Hybridization with TaylorF2 is used for the astrophysically uninteresting low frequency part for higher mass-ratios.

\begin{figure}[htbp]
  \centering
    \includegraphics[width=.45\textwidth]{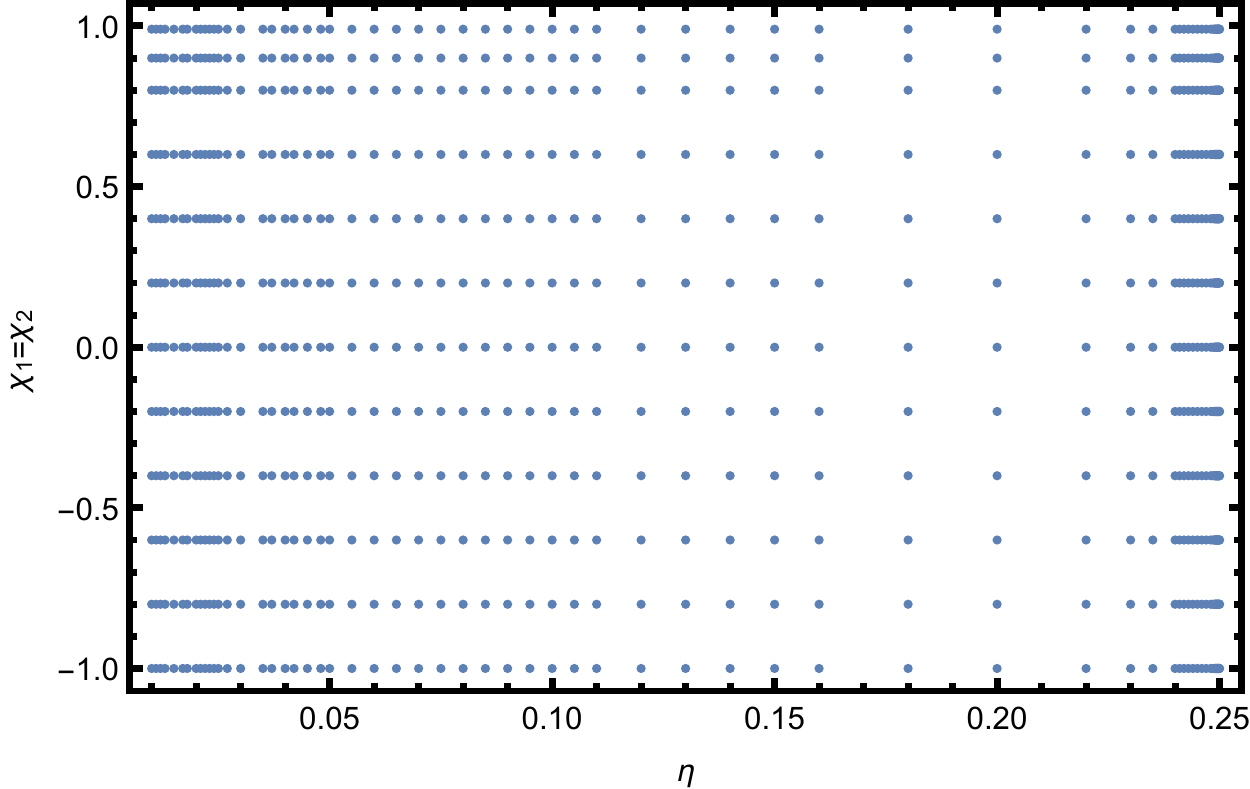}
  \caption{The location of input waveforms in $\eta$ and one of the aligned spins for the low frequency ROM. Both spin components use the same grid.}
  \label{fig:plots_grid_core_eta_chi1}
\end{figure}

The set of input waveforms was defined on a $67 \times 12 \times 12$ grid in $\eta, \chi_1, \chi_2$ (see Fig.~\ref{fig:plots_grid_core_eta_chi1}). The choice of points was obtained iteratively after checking the accuracy of the resulting ROM against SEOBNRv2 and refining the grid where needed. The sparse frequency grid used in the ROM covers $Mf \in [0.0000985, 0.15]$ using 200 points for the amplitude and 250 for the phase.

It should be pointed out that these waveforms are \emph{very} expensive to produce. The generation time is about 1.5 hours per waveform on an Intel Xeon Haswell with 2.5GHz clock speed. Depending on the configuration the memory required per waveform can reach up to $18$ GB towards the end of the EOB evolution. The $\sim 10000$ waveforms were generated on an MPI cluster with $4$ GB per core by filling the nodes only up to a third. The total cost was quite reasonable at about $50000$ core hours.

\subsection{High frequency ROMs} 
\label{sub:high_mass_high_chi1_rom}


At high frequency I use domain decomposition in the spin of the larger BH. The resolution requirements turned out to be more demanding for high $\chi_1$ (see Sec.~\ref{sub:non_smoothness_in_seobnrv2_at_for_high_spin}) and I therefore use a much finer grid for $\chi_1 \geq 0.4$ than for lower values of the spin. I describe both high frequency ROMs below, focusing on the high spin region.

Tests with a ROM covering the entire SEOBNRv2 parameter space showed unresolved structure for high spin on the larger BH at high total mass. To make a better model of this region it is efficient to work with a dense grid of short waveforms encompassing the late inspiral and merger ringdown. This model can then be combined with the low frequency ROM described in Sec.~\ref{sub:low_mass_rom}.

\begin{figure*}[htbp]
  \centering
    \includegraphics[width=.3\textwidth]{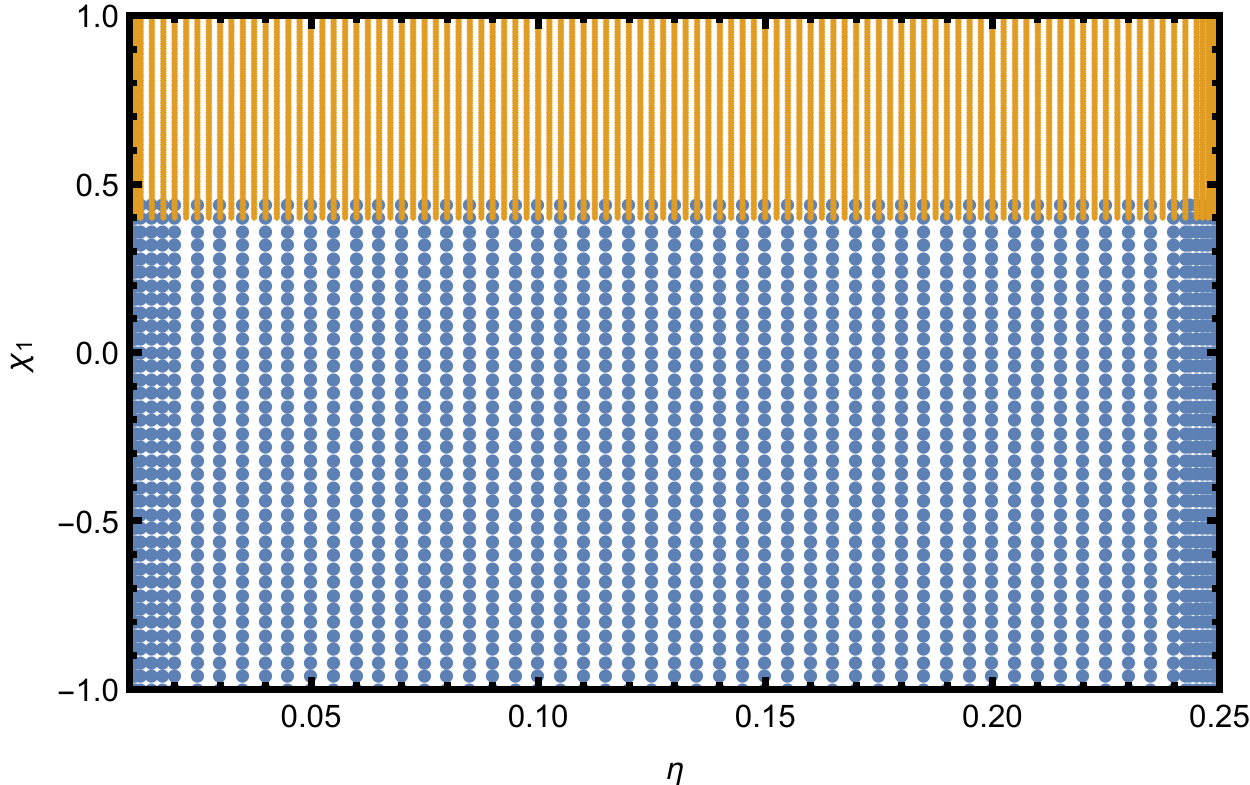}
    \includegraphics[width=.3\textwidth]{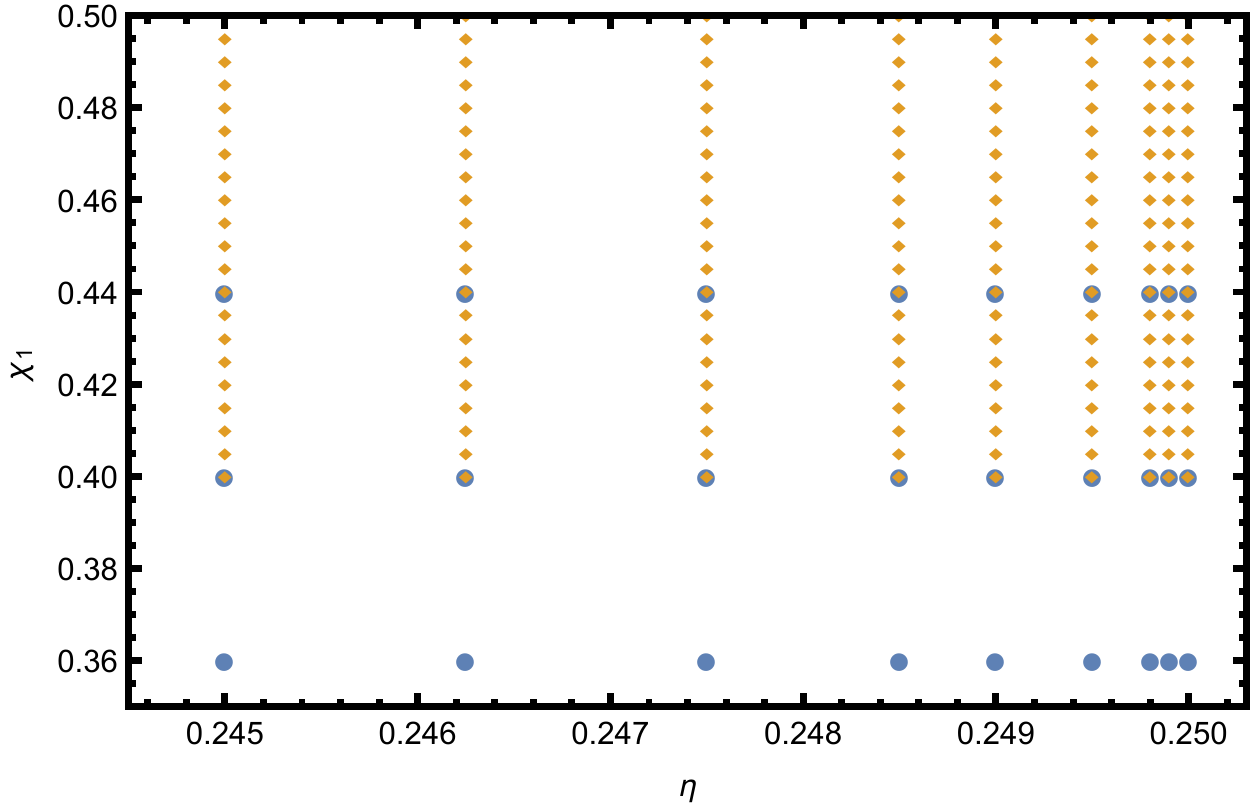}
    \includegraphics[width=.3\textwidth]{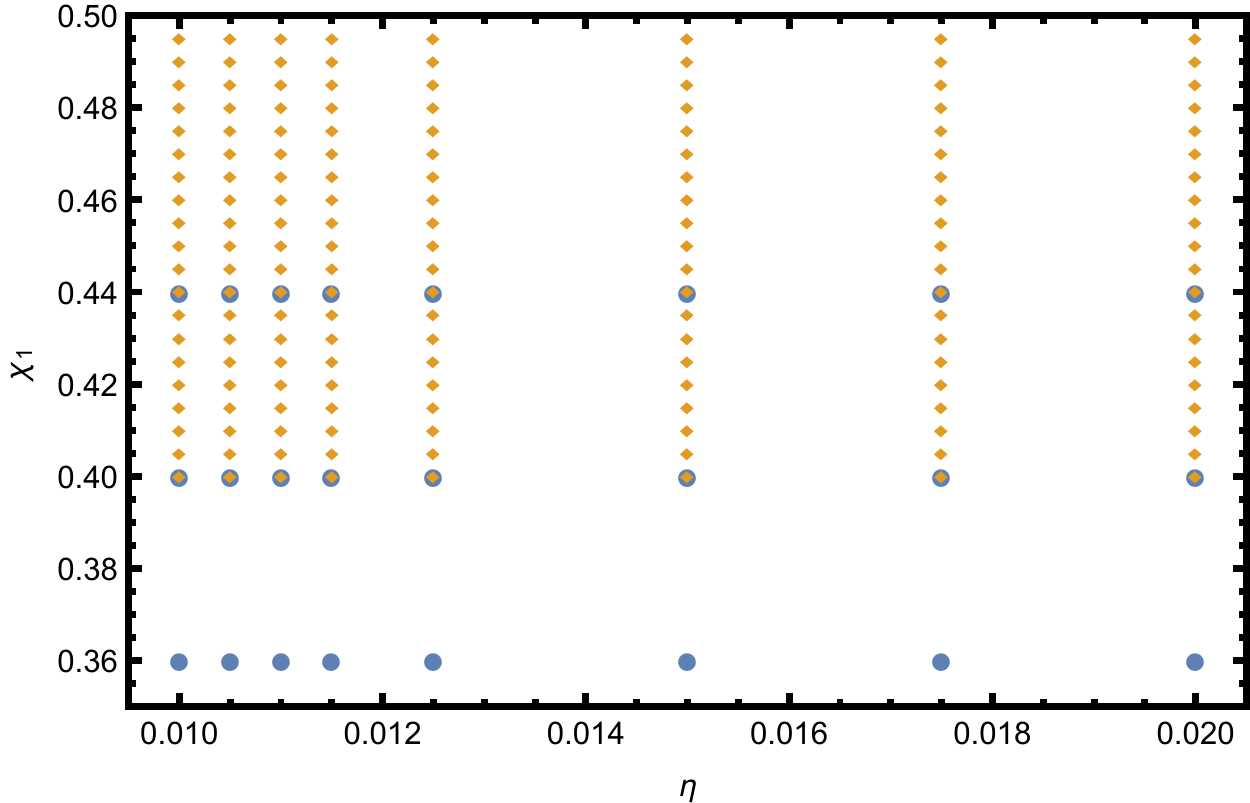}
  \caption{The grid of input waveforms in the symmetric mass-ratio $\eta$ and the aligned spin on the larger BH $\chi_1$ for the high frequency ROMs. The points for the ROM on the low $\chi_1$ patch are shown in blue and those for the ROM on the high $\chi_1$ patch in orange. The left panel shows the full domain, while the middle and right panel show the transition between low and high $\chi_1$ grids near equal mass and near the extreme mass-ratio boundary $\eta=0.01$ of the domain.}
  \label{fig:plots_grid_HL_eta_chi1}
\end{figure*}

After some refinements I chose a grid of $106 \times 119 \times 51 = 643314$ waveforms in $(\eta, \chi_1, \chi_2)$ covering
$\eta \in [0.01, 0.25]$ with $\Delta\eta \approx 0.0025$, 
$\chi_1 \in [0.4, 0.98999]$ with $\Delta\chi_1 \approx 0.005$, and
$\chi_2 \in [-1, 0.98999]$ with $\Delta\chi_2 \approx 0.04$. 
For the spins, the spacings were adjusted at the boundary, to conform to the range where SEOBNRv2 waveforms could be generated reliably. (The actual allowed range is $[-1, 0.99]$, but the LAL code~\cite{LAL-web} sometimes failed at these boundaries.) To improve accuracy near the $\eta$ boundaries several more points were added. 
The orange points in Fig.~\ref{fig:plots_grid_HL_eta_chi1} show the grid in $\eta$ and the spin on the larger BH, $chi_1$.

For $\chi_1 < 0.4$ a lower density in the spin $\chi_1$ turned out to be sufficient for good accuracy. Fine spacing near mass-ratio boundaries was still needed. I used a $63 \times 37 \times 51$ grid in $\eta, \chi_1, \chi_2$. The grid in $\chi_2$ is the same for the low and high $\chi_1$ ROM. The $\chi_1$ and $\chi_2$ grids are identical up to $\chi_1 = 0.4$. The blue points in Fig.~\ref{fig:plots_grid_HL_eta_chi1} show the grid in $\eta$ and $chi_1$ for this region.

The SEOBNRv2 waveforms were generated at a total mass of $100 M_\odot$ and were used in the frequency range $Mf \in [0.0074, 0.3]$ with $113$ sparse frequency points for the amplitude and phase. Compared to the input waveforms for the low frequency ROM (see Sec.~\ref{sub:low_mass_rom}) the computational cost for producing these high mass waveforms was small.

\section{Speedup and Accuracy} 
\label{sec:speedup_and_accuracy}

I present results on the performance of the overall ROM as defined in Sec.~\ref{sec:anatomy_of_the_seobnrv2_rom} in terms of the low frequency ROM and the two ROMs at high frequency. The main criteria for a successful ROM are that it facilities data analysis applications that were infeasible with the original model and that it is accurate in terms of the match or ``faithfulness'' between the ROM and the original model.

The speedup relative to SEOBNRv2 is summarized in Sec.~\ref{sub:speedup} and the faithfulness mismatch 
defined in Sec.~\ref{sub:inner_product_and_unfaithfulness}. Non-smooth features in SEOBNRv2 at high frequency are discussed in~\ref{sub:non_smoothness_in_seobnrv2_at_for_high_spin}. Finally, I discuss the faithfulness of the ROM against SEOBNRv2 in Sec.\ref{sub:unfaithfulness_against_seobnrv2} for second and third generation ground-based GW detectors and a wide range of total masses.

\subsection{Speedup} 
\label{sub:speedup}

\begin{figure}[h]
  \centering
    \includegraphics[width=.49\textwidth]{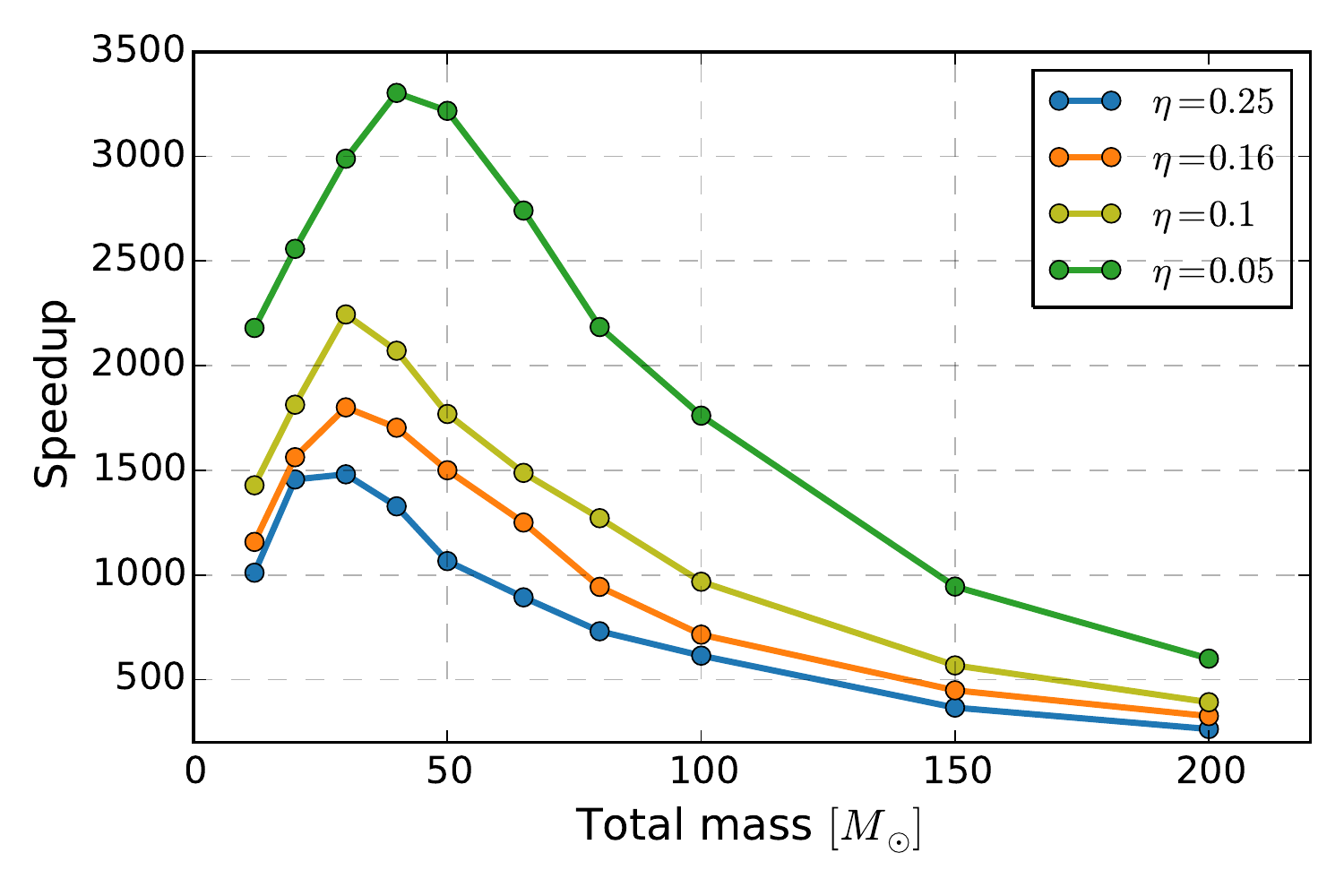}
  \caption{Speedup of waveform generation of the overall ROM compared to SEOBNRv2.}
  \label{fig:plots_Speedup_v2DS_HI_mass_etas}
\end{figure}

The speedup of the overall ROM as implemented in LAL~\cite{LAL-web} against SEOBNRv2 is shown in Fig.~\ref{fig:plots_Speedup_v2DS_HI_mass_etas}. The full number of frequency points obtained from the FFT of a time-domain waveform sampled at a rate of $16384$ Hz is used. The speedup depends both on the total mass and the mass-ratio. The figure only shows the speedup down to a total mass of $12 M_\odot$; for lower total mass the generation of high mass-ratio SEOBNRv2 waveforms becomes prohibitively expensive.
As discussed in ~\cite{Purrer:2014fza} the evaluation speed of the ROM is composed of a constant cost per frequency point for the spline interpolation in frequency and a startup cost for performing the interpolation of coefficients over the parameter space. The former dominates for long waveforms while the latter only becomes significant for very short waveforms. The speedup peaks around $20 - 50M_\odot$ reaching a factor of several thousands. For high total mass it falls off to several hundreds.

Due to gluing the ROM in frequency parameter space interpolation needs to be carried out in both frequency patches which leads to about a factor of two slowdown compared to a single-patch in frequency ROM.
The speedup does not significantly depend on whether the configuration is in the low or high $\chi_1$ patch, even though the data set is much larger for high spin. This is due to the local support of the cubic splines used to interpolate the amplitude and phase coefficients. The larger size of the high $\chi_1$ dataset only impacts the initial step where the ROM data is loaded into memory. Since this happens only once for each analysis it does not affect performance.

Further speedup can be obtained by truncating of the expansion in amplitude and phase used in the ROM to $\sim 50$ SVD modes (see Eq. (6.3) in~\cite{Purrer:2014fza}). This yields an additional speedup factor of $3$ for high total mass and also reduces the size of the data set. This increase in speed comes at the cost of higher mismatch errors near the high mass-ratio and extreme anti-aligned spin boundaries that can reach up to $0.3\%$.

Some data analysis applications require a significantly smaller number of frequency points, e.g. the computation of the match in zero noise with a detector noise curve, and thus the speedup can be \emph{much} larger than the conservative numbers shown in Fig.~\ref{fig:plots_Speedup_v2DS_HI_mass_etas}. For example parameter estimation using a match-based likelihood~\cite{Puerrer-spin-PE} and the generation of template banks with stochastic algorithms~\cite{Harry:2009ea,Manca:2009xw} can use fairly coarse frequency spacing $O(Hz)$ without compromising the accuracy of the match computation.

Similarly, the ROM can be exploited for the construction of reduced order quadratures~\cite{Canizares:2013ywa,Canizares:2014fya} to provide further speedup in calculating inner products for low mass systems.

\subsection{Inner product and unfaithfulness} 
\label{sub:inner_product_and_unfaithfulness}

A natural inner product between two waveforms $h_1, h_2$ can be defined as
\begin{equation}
  \label{eq:inner_product}
  \langle h_1, h_2 \rangle := 4 \mathrm{Re} \int_{f_\mathrm{low}}^{f_\mathrm{high}} \frac{\tilde h_1(f) \tilde h_2^*(f)}{S_n(f)} df,
\end{equation}
where $\tilde{}$ denotes Fourier transformation, $S_n(f)$ is the power spectral density (PSD) of the detector noise and $f_\mathrm{low}, f_\mathrm{high}$ are suitable cutoff frequencies for the detector sensitivity. The lower cutoff depends on the PSD I choose below while I keep the higher cutoff fixed at $8000$ Hz.
The mismatch (or ``unfaithfulness'') is defined as the normalized inner product~\eqref{eq:inner_product} maximized over time and phase shifts
\begin{equation}
  \label{eq:mismatch}
  \mathcal{M} := 1 - \max_{t_0, \phi_0} \frac{\langle h_{22}^\mathrm{SEOBNRv2}, h_{22}^\mathrm{ROM} \rangle}{\Vert h_{22}^\mathrm{SEOBNRv2} \rVert \, \lVert h_{22}^\mathrm{ROM} \rVert},
\end{equation}
where $\Vert h \rVert := \sqrt{\langle h, h \rangle}$.


\subsection{Non-smoothness in SEOBNRv2 at for high spin} 
\label{sub:non_smoothness_in_seobnrv2_at_for_high_spin}

Before discussing the general accuracy of the ROM in terms of unfaithfulness in Sec.~\ref{sub:unfaithfulness_against_seobnrv2}, I first focus on the high frequency region where the detectors are sensitive to the merger-ringdown of BH binary coalescences and point out an unexpected feature in SEOBNRv2.

I compute the faithfulness against 8000 SEOBNRv2 test waveforms at a total mass of $150 M_\odot$ for aLIGO design with $10$ Hz starting frequency. The mismatches are smaller than $0.1\%$ everywhere except along specific lines in the parameter space.

\begin{figure}[htbp]
  \centering
    \includegraphics[width=.5\textwidth]{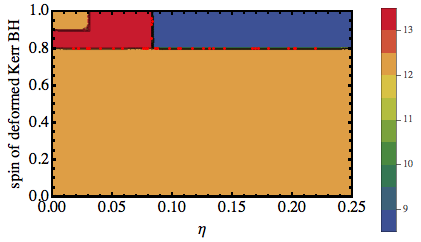}
  \caption{The ``comb size'' parameter used in SEOBNRv2 is discontinuous over the parameter space as indicated by the contours. Superimposed are configurations with mismatches $> 0.1\%$ as shown in \ref{fig:plots_Faith_h100_chi1_eta}. They lie exactly on lines $\chi = 0.8$ and $\eta = 10/121 \approx 0.083$ where the comb size has a discontinuity.}
  \label{fig:plots_comb_size}
\end{figure}

Fig.~\ref{fig:plots_comb_size} shows a contour plot of the ``comb size'' parameter over which the inspiral-plunge and merger-ringdown waveforms are matched in SEOBNRv2~\cite{ATcomment,Taracchini:2013rva} as a function of symmetric mass-ratio $\eta$ and the spin of the deformed background Kerr BH $\chi$ (see Eqn.~\eqref{eq:chi_bg}). The comb size has a discontinuity at $\chi = 0.8$ and $\eta = 10/121$ and along these lines the mismatch of test waveforms turns out to be worse than $0.1\%$. These configurations are shown as red dots in Fig.~\ref{fig:plots_comb_size}.

\begin{figure}[htbp]
  \centering
    \includegraphics[width=.49\textwidth]{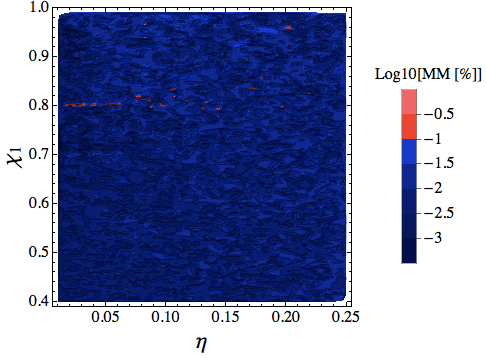}
  \caption{Faithfulness mismatches (MM) over the parameter space as a function of the spin of the larger BH $\chi_1$ and the symmetric mass-ratio $\eta$ and as a function of the spin of the deformed background Kerr BH $\chi$ used in the EOB model.}
  \label{fig:plots_Faith_h100_chi1_eta}
\end{figure}

In terms of the spin on the large BH $\chi_1$ the discontinuity appears as a line for high mass-ratios, but for smaller mass-ratios where $\chi_2$ becomes more important, there is significant scatter. This is shown in Fig.~\ref{fig:plots_Faith_h100_chi1_eta}.

Non-smooth features such as the above would preclude spectral interpolation from working efficiently. Spline interpolation is more robust and the effect of non-smoothness is confined due to localized support of the basis functions.

\subsection{Unfaithfulness against SEOBNRv2} 
\label{sub:unfaithfulness_against_seobnrv2}

\begin{figure*}[htbp]
  \centering
    \includegraphics[width=.49\textwidth]{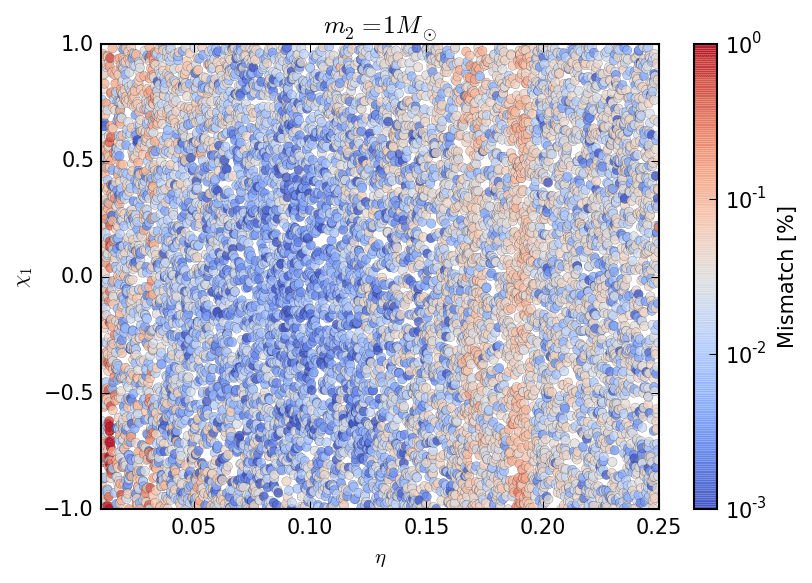}
    \includegraphics[width=.49\textwidth]{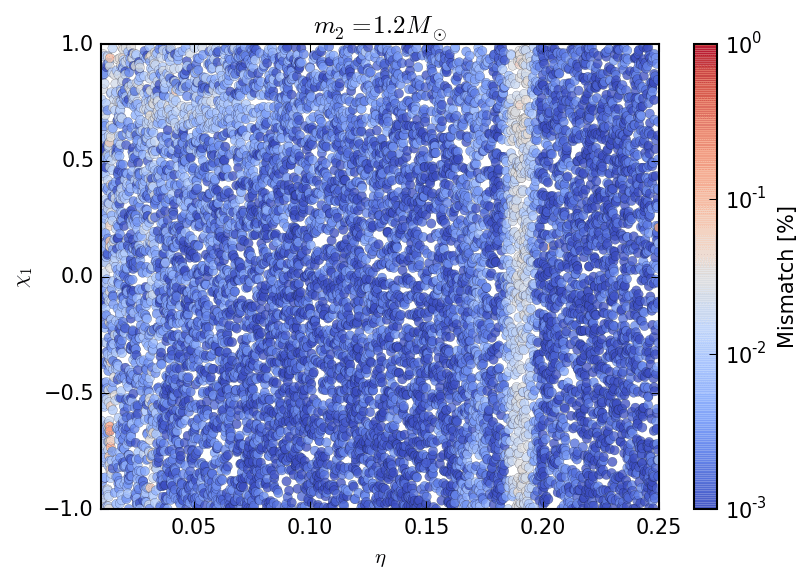}
    \includegraphics[width=.49\textwidth]{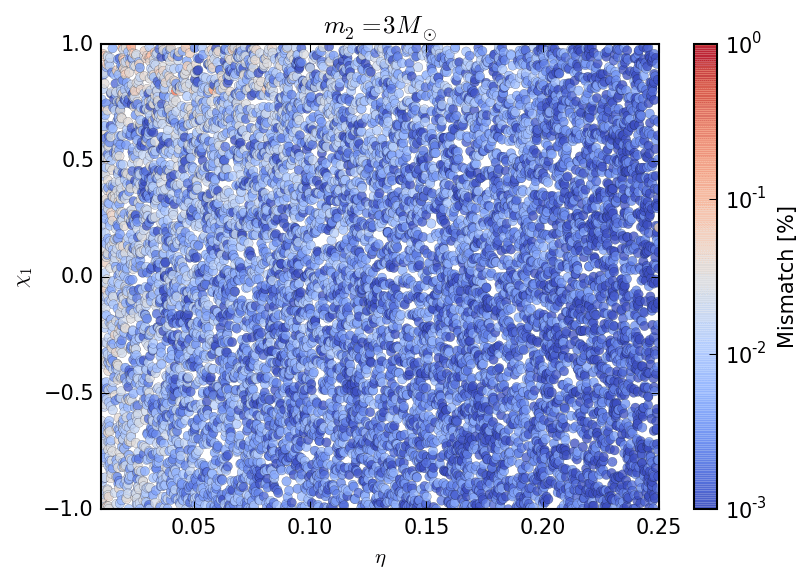}
    \includegraphics[width=.49\textwidth]{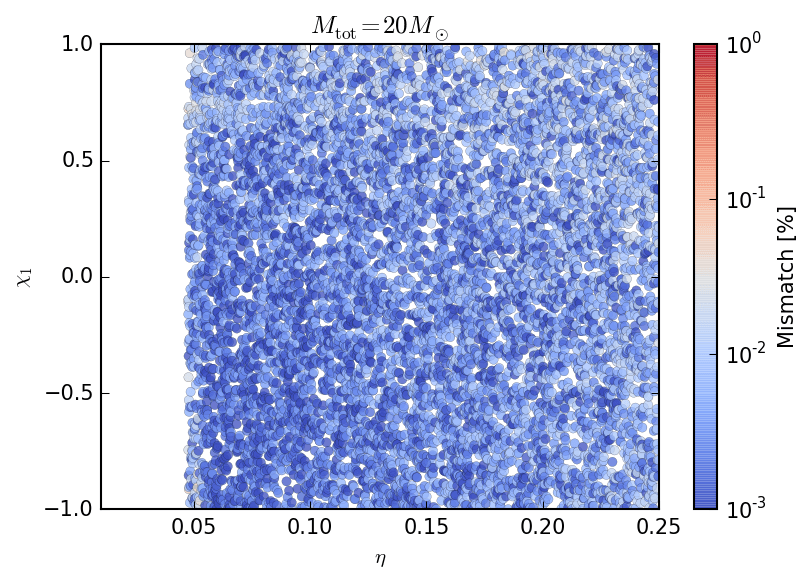}
    \includegraphics[width=.49\textwidth]{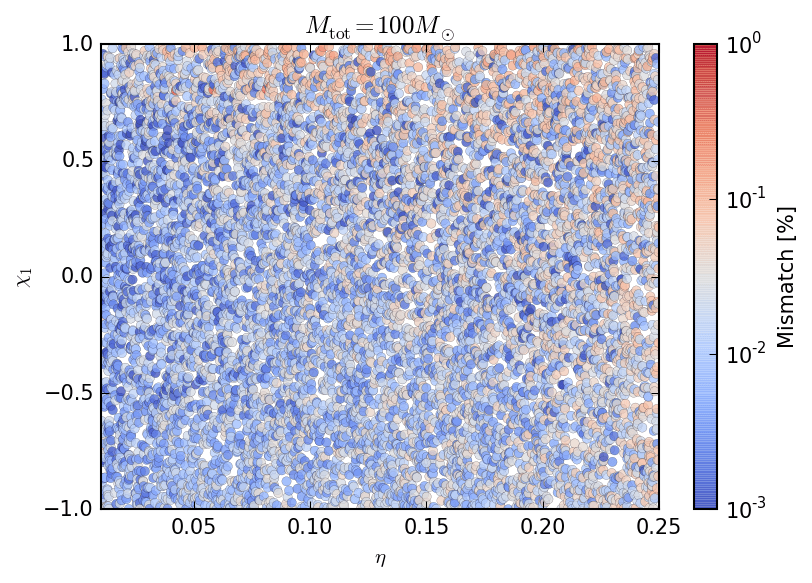}
    \includegraphics[width=.49\textwidth]{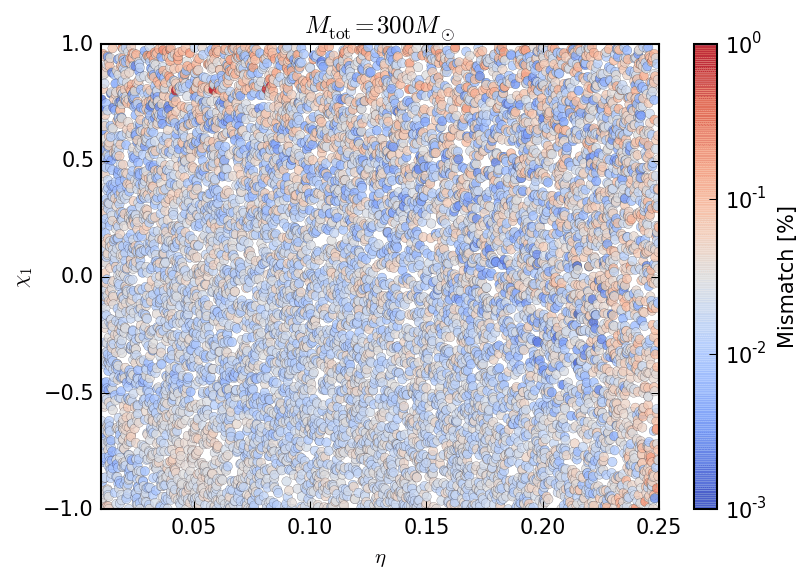}
  \caption{Mismatches between the overall ROM and SEOBNRv2 for aLIGO design with a low frequency cutoff of $10$ Hz. To facilitate comparisons between panels, the colorbar scales are fixed at a mismatch of $1\%$. The worst mismatches at high mass lie on lines where SEOBNRv2 is non-smooth (see Sec.~\ref{sub:non_smoothness_in_seobnrv2_at_for_high_spin}). Histograms of the mismatches are shown in Fig.~\ref{fig:plots_glued_ROM_DSv2_M2_Histograms} (left panel).
}
  \label{fig:MM_scatter_aLIGO_zdethp_10Hz}
\end{figure*}

\begin{figure*}[htbp]
  \centering
    \includegraphics[width=.49\textwidth]{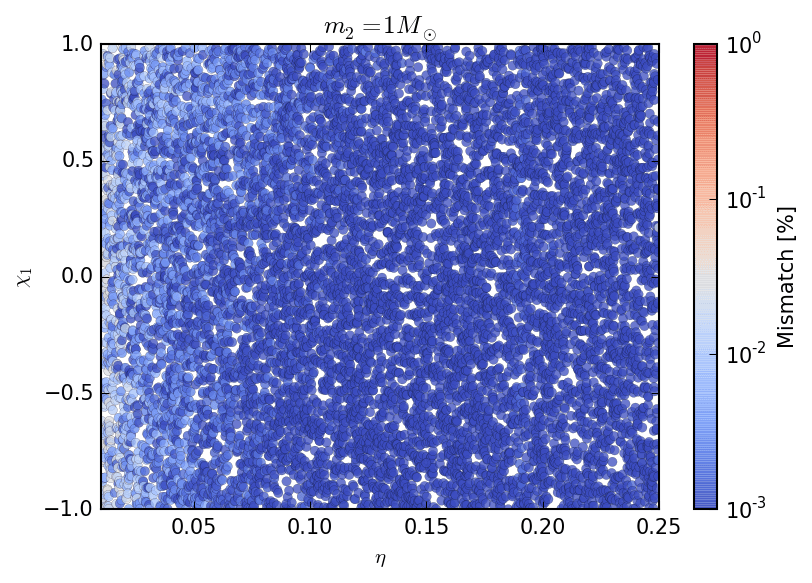}
    \includegraphics[width=.49\textwidth]{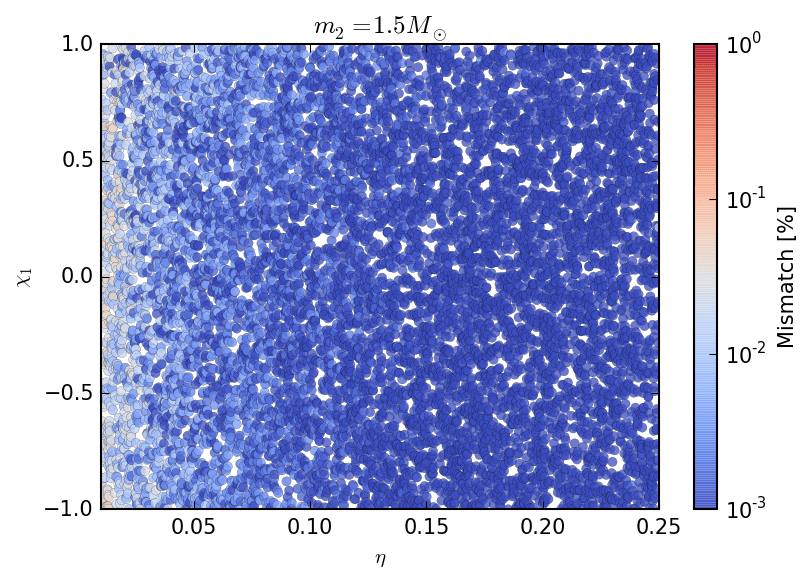}
    \includegraphics[width=.49\textwidth]{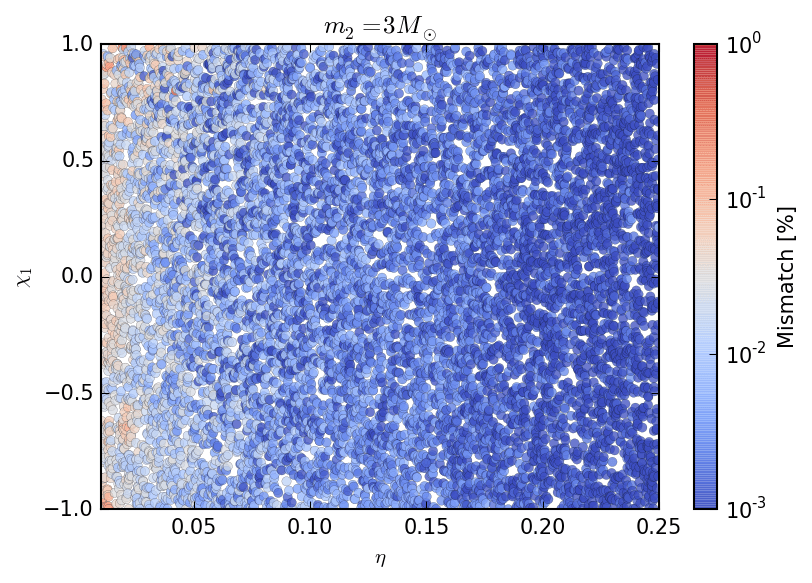}
    \includegraphics[width=.49\textwidth]{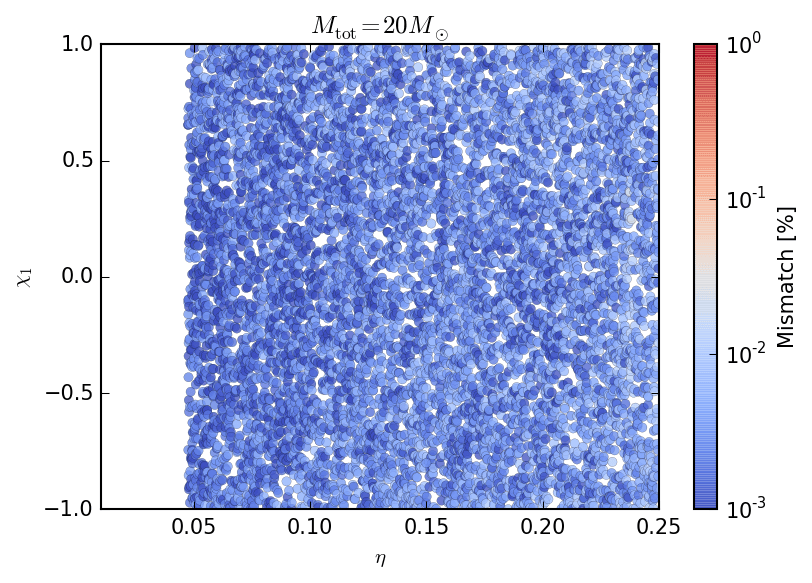}
    \includegraphics[width=.49\textwidth]{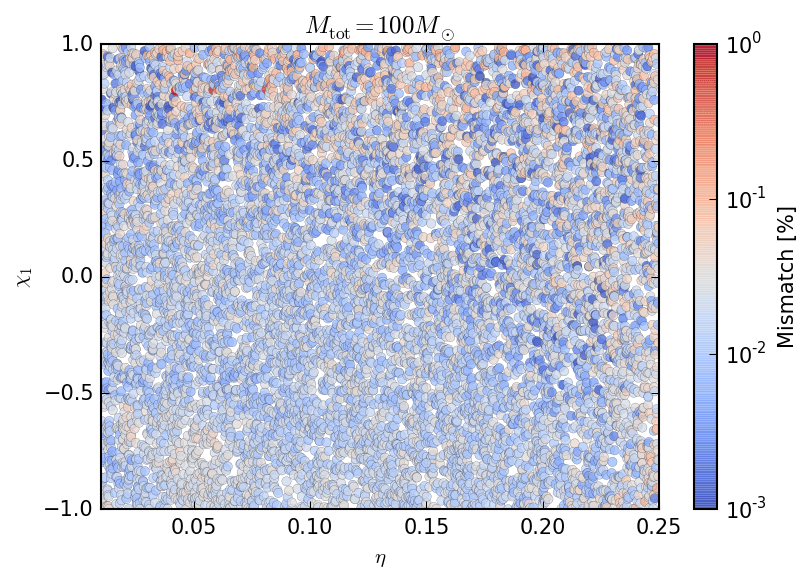}
    \includegraphics[width=.49\textwidth]{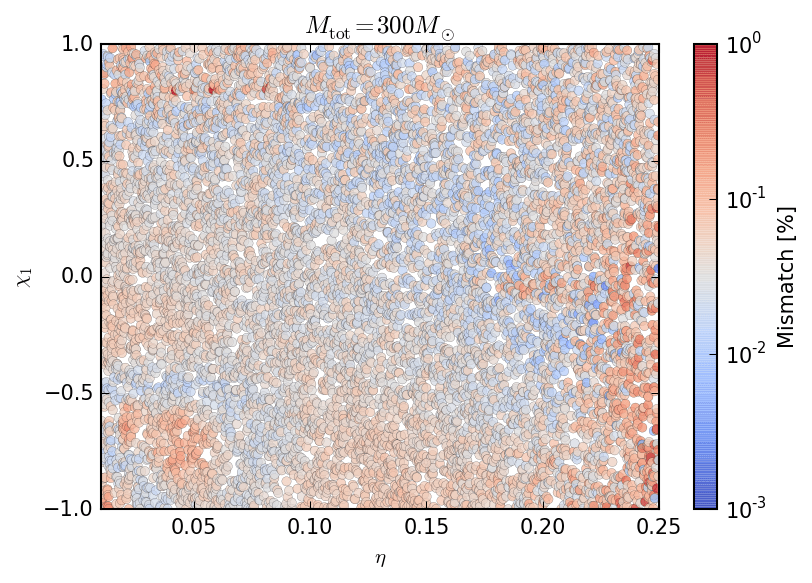}
  \caption{Mismatches between the overall ROM and SEOBNRv2 for early aLIGO with a low frequency cutoff of $30$ Hz. To facilitate comparisons between panels, the colorbar scales are fixed at a mismatch of $1\%$.
  As in Fig.~\ref{fig:MM_scatter_aLIGO_zdethp_10Hz} the worst mismatches at high mass lie on lines where SEOBNRv2 is non-smooth (see Sec.~\ref{sub:non_smoothness_in_seobnrv2_at_for_high_spin}). Histograms of the mismatches are shown in Fig.~\ref{fig:plots_glued_ROM_DSv2_M2_Histograms} (right panel).
  }
  \label{fig:MM_scatter_aLIGO_early_30Hz}
\end{figure*}

Here I discuss the accuracy of thr ROM in terms of the faithfulness mismatch~\eqref{eq:mismatch} between the ROM and SEOBNRv2, i.e. the overlap is only optimized over time and phase while mass-ratio and spins are fixed. The results were obtained from about 10000 random configurations with parameters $\eta, \chi_1, \chi_2$ uniformly distributed in the allowed domain. The results for aLIGO are presented in Fig.~\ref{fig:MM_scatter_aLIGO_zdethp_10Hz} for the design PSD and Fig.~\ref{fig:MM_scatter_aLIGO_early_30Hz} for the early aLIGO PSD and in table~\ref{tab:mismatches} where additional data points are given as well.

Fig.~\ref{fig:MM_scatter_aLIGO_zdethp_10Hz} shows mismatches (in percent) for the aLIGO design (zero-detuned high power) PSD~\cite{T0900288} with a lower frequency cutoff of $10$ Hz. In the first three panels the mass of the smaller body $m_2$ is fixed while the symmetric mass-ratio covers the entire domain $0.01 \leq \eta \leq 0.25$. This is done to restrict to relevant astrophysical systems with a smallest component mass above $1 M_\odot$. For $m_2 = 1 M_\odot$ the mismatch is in general around $\sim 0.5\%$ or lower, except for some configurations at very high mass-ratios where the mismatch can reach up to $\sim 1.5\%$. The mismatches improve considerably if the smallest component mass is increased to be $m_2 = 1.2M_\odot$ and $3 M_\odot$. There, the median of the mismatch distribution is $\sim 0.003\%$ and the worst mismatch $0.25\%$.

\begin{figure*}[htbp]
  \centering
  \includegraphics[width=.49\textwidth]{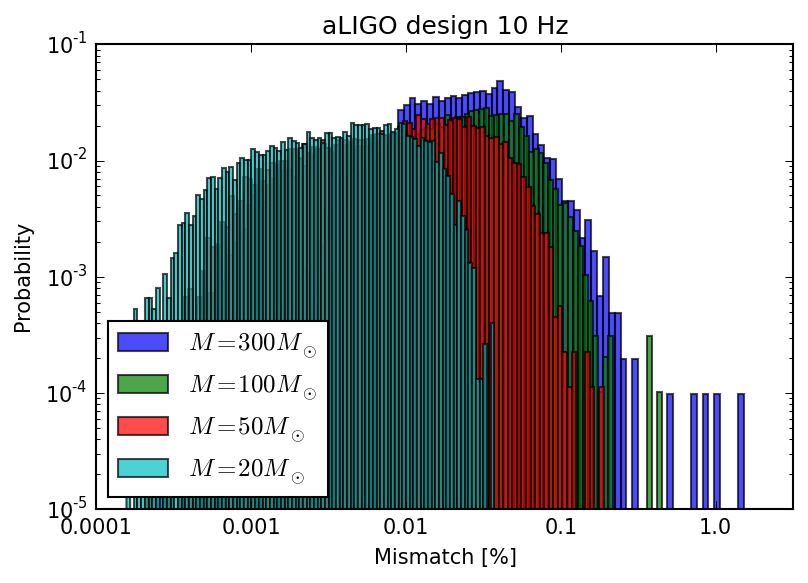}
    \includegraphics[width=.49\textwidth]{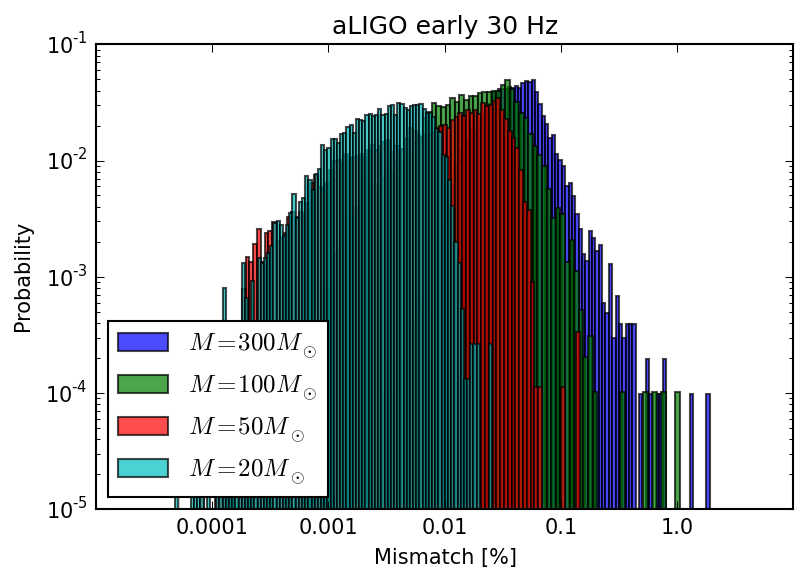}
  \caption{Mismatch histograms for aLIGO zdethp $10$ Hz (left) and aLIGO early $30$ Hz (right). At high total mass isolated points have mismatch close to (or above) $1\%$. These configurations are due to non-smoothness in SEOBNRv2.
  }
  \label{fig:plots_glued_ROM_DSv2_M2_Histograms}
\end{figure*}

In the last three panels of Fig.~\ref{fig:MM_scatter_aLIGO_zdethp_10Hz} the total mass is kept constant. For $M_\mathrm{tot} = 20 M_\odot$ the model only covers systems up to mass-ratio $q=19$ or $\eta = 0.0475$. Beyond this value no mismatches are computed so that the mass of the light body is at least $1 M_\odot$. 
The median mismatch is below $0.005\%$ for $M_\mathrm{tot} = 20 M_\odot$ and rises to $0.02\%$ for $M_\mathrm{tot} = 300 M_\odot$. The worst mismatch also rises with the total mass as the merger ringdown is shifted to lower frequencies in the detector sensitivity curve. At $M_\mathrm{tot} = 20 M_\odot$ the worst mismatch is below $0.06\%$. It rises to $0.4\%$ and $1.5\%$ for $100$ and $300 M_\odot$. As can also be seen from the histogram in the left panel of Fig.~\ref{fig:plots_glued_ROM_DSv2_M2_Histograms} the mismatch is higher than the continuum for a few configurations only. These configurations lie close to the lines where the SEOBNRv2 comb size parameter has a discontinuity and the waveform does not depend smoothly on parameters (see Sec.~\ref{sub:non_smoothness_in_seobnrv2_at_for_high_spin}). After removing these points, the worst mismatch at high total mass is $\sim 0.3\%$.

Analogous to the results for aLIGO design in Fig.~\ref{fig:MM_scatter_aLIGO_zdethp_10Hz}
I show mismatches for the early aLIGO PSD~\cite{G1000176} with a lower frequency cutoff of $30$ Hz in Fig.~\ref{fig:MM_scatter_aLIGO_early_30Hz}.
The accuracy at low masses is even better for the early aLIGO noise curve. The median mismatch is below $0.003\%$ and the worst mismatch is below $0.1\%$ except at very high mass-ratio where it can reach $0.4\%$ for $M_\mathrm{tot} = 3 M_\odot$.
As the total mass increases the median mismatch goes from $0.003\%$ at $20 M_\odot$ to $0.04\%$ at $300 M_\odot$. The worst mismatch rises from $0.03\%$ to $1\%$ and $1.9\%$ at $20, 100, 300 M_\odot$. The worst mismatches are again due to the non-smoothness of SEOBNRv2. After removing the points very close to where the comb size parameter is discontinous the worst mismatch becomes $0.2\%$ and $0.8\%$ at $100, 300 M_\odot$. Compared to the results for aLIGO design there is some degradation of the accuracy for very high total masses. This is mostly due to the different shape of the early aLIGO PSD which tends to highlight waveform disagreement at high frequencies.

\begin{figure*}[htbp]
  \centering
    \includegraphics[width=.49\textwidth]{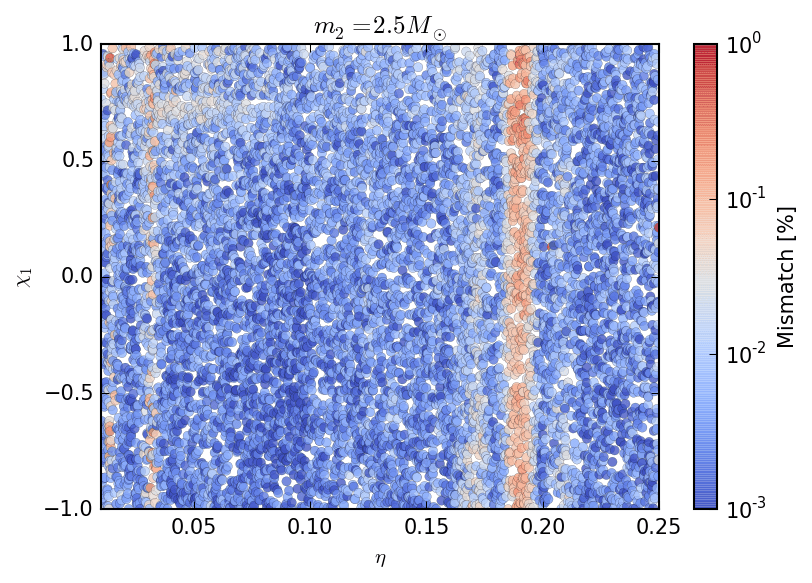}
    \includegraphics[width=.49\textwidth]{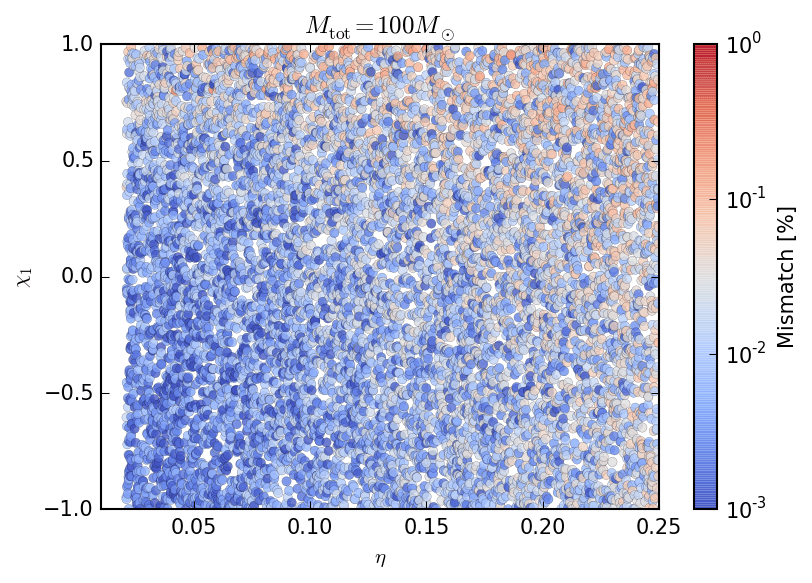}
  \caption{Mismatches between the (glued) ROM and SEOBNRv2 for ET-D with a low frequency cutoff of $5$ Hz.}
  \label{fig:MM_scatter_ETD_5Hz}
\end{figure*}

\begin{figure*}[htbp]
  \centering
    \includegraphics[width=.49\textwidth]{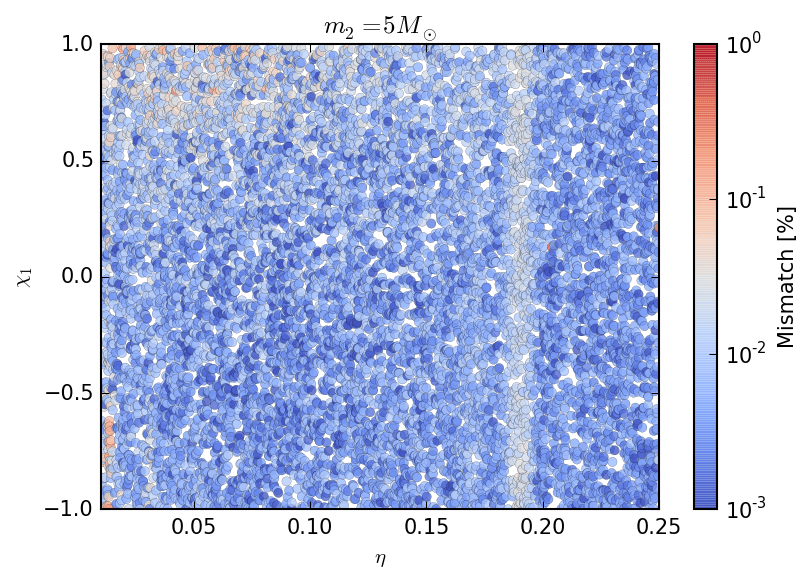}
    \includegraphics[width=.49\textwidth]{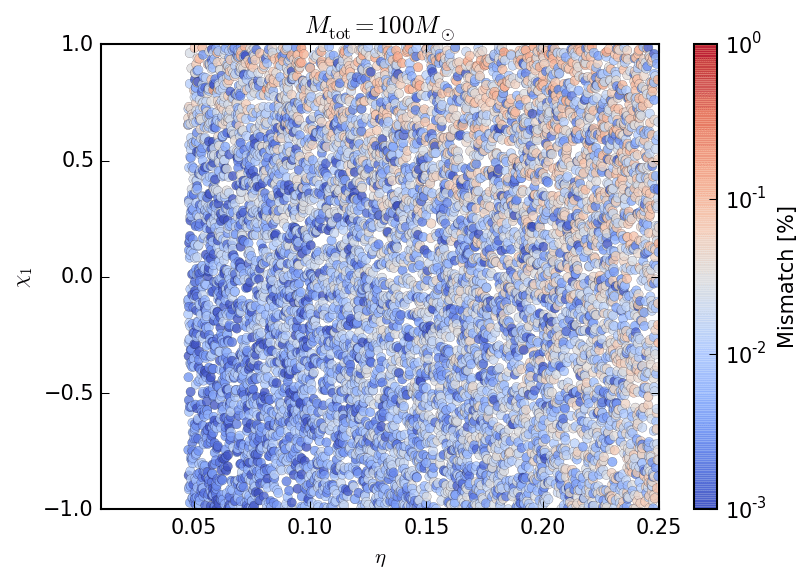}
  \caption{Mismatches between the (glued) ROM and SEOBNRv2 for ET-D with a low frequency cutoff of $2$ Hz.}
  \label{fig:MM_scatter_ETD_2Hz}
\end{figure*}

Finally, I compute mismatches using the ET-D noisecurve~\cite{ET-D} for the third generation Einstein Telescope~\cite{Punturo:2010zza,Sathyaprakash:2012jk}. Since the ROM is defined on a grid in geometric frequency $Mf$, the low frequency cutoff can be decreased by simultaneously increasing the total mass by the same factor. It is then possible to use the ROM for systems with a lighter component larger than $2M_\odot$ at a $5$ Hz lower cutoff (see Fig.~\ref{fig:MM_scatter_ETD_5Hz}) or $5M_\odot$ at a $2$ Hz lower cutoff (see Fig.~\ref{fig:MM_scatter_ETD_2Hz}). In the left panel of Fig.~\ref{fig:MM_scatter_ETD_5Hz} the component mass was increased to $2.5 M_\odot$. Then the worst mismatch is below $\sim 0.6\%$ with the median at $0.005\%$. The low mass accuracy in the left panel of Fig.~\ref{fig:MM_scatter_ETD_2Hz} with the $2$ Hz lower cutoff is comparable. The accuracy at high mass is also comparable between the two choices of lower frequency cutoff. The worst mismatch is below $\sim 0.3\%$ with the median at $0.01\%$.

\begin{table*}
\begin{tabular}{l|lll|llllll}
  \hline
  \hline
 Unfaithfulness & $m_2 = 1 M_\odot$ & $1.2 M_\odot$ & $3 M_\odot$ & $M = 20 M_\odot$  & $50 M_\odot$ & $80 M_\odot$ & $100 M_\odot$ & $200 M_\odot$ & $300 M_\odot$\\
  \hline
  \hline
  \multicolumn{10}{l}{aLIGO design 10 Hz}\\
  \hline
  $max_\lambda \mathcal{M} [\%]$ & $1.5 \, (0.03 \, \text{for} \, \eta>0.05)$  & $0.22$   & $0.27$ & $0.06$  & $0.19$  & $0.53 \, (0.3)$ & $0.45 \, (0.3)$  & $1.0 \, (0.26)$ & $1.5 \, (0.3)$ \\
  $med_\lambda \mathcal{M} [\%]$ & $0.02$                    & $0.002$ & $0.003$ & $0.004$ & $0.009$ & $0.14$ & $0.016$ & $0.02$       & $0.024$ \\
  \hline
  \hline
  \multicolumn{10}{l}{aLIGO design 20 Hz}\\
  \hline
  $max_\lambda \mathcal{M} [\%]$ & $0.23$  & $0.11$  & $0.29$  & $0.03$  & $0.18$  & $0.34 \, (0.2)$ & $0.47 \, (0.2)$  & $0.9 \, (0.25)$ & $1.37 \, (0.3)$ \\
  $med_\lambda \mathcal{M} [\%]$ & $0.001$ & $0.001$ & $0.003$ & $0.004$ & $0.009$ & $0.14$ & $0.016$ & $0.02$       & $0.024$ \\
  \hline
  \hline
  \multicolumn{10}{l}{aLIGO early 30 Hz}\\
  \hline
  $max_\lambda \mathcal{M} [\%]$ & $0.05$   & $0.08$   & $0.38$  & $0.026$ & $0.15$ & $0.63 \, (0.13)$  & $1.1 \, (0.18)$ & $1.6 \, (0.4)$ & $1.9 \, (0.76)$ \\
  $med_\lambda \mathcal{M} [\%]$ & $0.0006$ & $0.0007$ & $0.003$ & $0.003$ & $0.01$ & $0.015$        & $0.018$ & $0.029$ & $0.04$  \\
  \hline
  \hline
\end{tabular}
\caption{
    \label{tab:mismatches}
        Faithfulness mismatches of the ROM against SEOBNRv2 for alIGO design with a lower cutoff frequency of $10$ or $20$ Hz and early aLIGO with $30$ Hz. The mismatch is given for a range of fixed masses of the smaller body $m_2$ and for a range of fixed total masses $M$. I quote both the maximum and the median of the mismatch distribution in \%. For high total masses $M\geq 80 M_\odot$ I give the overall maximum mismatch and the maximum mismatch with points near $\chi \sim 0.8$ removed where SEOBNRv2 is non-smooth in parentheses.
    }
\end{table*}

\section{Conclusions} 
\label{sec:conclusions}

I have presented a reduced order (or surrogate) model of SEOBNRv2~\cite{Taracchini:2013rva} for the dominant harmonic of the GW emitted by BH binary systems. The ROM covers the entire domain of definition of SEOBNRv2: $0.01 \leq \eta \leq 0.25$ and $-1 \leq \chi_i \leq 0.99$. The ROM can be used for compact binaries from a total mass of $2 M_\odot$ and larger with a $10$ Hz lower frequency cutoff, or more generally for geometric frequencies from $Mf_\text{low} = 2 M_\odot[s] 10 \mathrm{Hz} = 0.0000985$ to $Mf_\text{high} = 0.3$.
This covers the entire mass range range of BNS, NS-BH and BBH sytems of interest for second generation ground-based laser-interferometric detectors aLIGO and AdvVIRGO. The ROM can also be use for data analysis with ET for slightly heavier systems from $\sim 2 - 5 M_\odot$ onwards depending on the lower frequency cutoff.

While this ROM is based on the method described in~\cite{Purrer:2014fza}, the construction relies on extensions to these techniques (see Secs.~\ref{sec:reduced_order_modeling_techniques} and~\ref{sec:anatomy_of_the_seobnrv2_rom}). In particular, the parameter space is decomposed into two 3-dimensional patches in $\eta, \chi_1, \chi_2$ and a division into a low and a high frequency patch in $Mf$. This domain decomposition technique makes it possible to model the smooth inspiral part with a coarser waveform set where waveform generation is most costly and to obtain higher resolution for the late inspiral and merger-ringdown, where waveform generation is much cheaper. The ringdown is fit against a Lorentzian to guarantee that the complete ringdown is modeled for all configurations in the SEOBNRv2 parameter space.

As shown in Fig.~\ref{fig:plots_Speedup_v2DS_HI_mass_etas} this ROM is several thousand times faster than SEOBNRv2 when using the full number of frequency points. This speedup is crucial for searches and parameter estimation in the LSC~\cite{LSC_web}. The SEOBNRv2 ROM has already been used in theoretical parameter estimations studies within the LSC~\cite{Veitch:2015ela,Haster:2015cnn} and is used in aLIGO's first observing run.
Data analysis applications such as the generation of stochastic template banks for searches~\cite{Harry:2009ea,Manca:2009xw,PhysRevD.89.084041,PhysRevD.89.024003,PhysRevD.89.024010} and parameter estimation studies in zero noise~\cite{Puerrer-spin-PE} that can elucidate important properties of waveform models require only the computation of matches with a PSD. For these applications speedups can be several tens of thousands.

Very recently an optimized version of the SEOBNRv2 code has become available~\cite{Etienne-v2-opt} in LALSuite~\cite{LAL-web}, although it has not yet been reviewed in the LSC. This code provides a speedup of one to two orders of magnitude over the reviewed SEOBNRv2, with the largest speedups observed for the longest waveforms (e.g., binary neutron stars). However, the ROM presented here is still more than an order of magnitude faster than this optimized code if all frequency points are used and can be a factor thousand faster for specific data analysis applications mentioned above. The availability of optimized EOB codes is very beneficial for the construction of ROMs.

A previous version of this ROM (\texttt{SEOBNRv2\_ROM\_DoubleSpin}) has been reviewed by the LSC and is used to build template banks for searches and as a model in parameter estimation simulations during the first observing run of aLIGO. An implementation of the ROM described in this paper is available under the name of \texttt{SEOBNRv2\_ROM\_DoubleSpin\_HI} in the LSC Algorithmic Library Suite (LALSuite)~\cite{LAL-web}. Compared to the previous version the ROM presented in this paper better resolves the region where SEOBNRv2 is non-smooth, introduces patching in frequency, and thereby significantly extends the total mass range from $12 M_\odot$ down to BNS systems with $2 M_\odot$.

Apart from the speedup, a ROM must provide a good approximation of the original waveform. This can be quantified by the match or the faithfulness between the two models over the parameter space.
The unfaithfulness for SEOBNRv2 against SXS waveforms was found to be less than $1\%$ for total masses from $20 M_\odot$ to $200 M_\odot$~\cite{Taracchini:2013rva}. In this mass range the mismatch between the ROM and SEOBNRv2 is always well below the SEOBNRv2 calibration accuracy except along lines where SEOBNRv2 has been shown to be non-smooth (see Sec.~\ref{sub:non_smoothness_in_seobnrv2_at_for_high_spin}) for high total masses.
Below $20 M_\odot$ the ROM agrees in general extremely well (mismatch $\lesssim 0.1\%$) with SEOBNRv2.
Only for extreme NS-BH systems with very high mass-ratio and a very light NS with a $10$ Hz lower cutoff does the mismatch reach $1\%$.

Beyond faithfulness GW searches and PE require the location of the amplitude maximum in the time domain strain is accurate. \ref{sub:chirp_time} describes how a time shift can be applied to the ROM phasing in order to achieve this. Moreover, useful functions can be defined for the length of a waveform in time as a function of a starting frequency and its inverse.

The techniques described here can be applied to building ROMs for aligned-spin waveform models including higher harmonics, with a mode-by-mode approach~\cite{Marsat-ROM}. Modeling GWs emitted from precessing binaries is a much harder problem to tackle due to the significantly larger size of the spin space and will require the development of further techniques.



\appendix
\section{Chirp Time and correction of time of coalescence} 
\label{sub:chirp_time}

A ``time of frequency'' function and its inverse can be defined from the Fourier domain phasing. These functions are useful when the template duration or ``chirp time''~\cite{PhysRevD.50.R7111} is needed by data analysis algorithms and also allow to correct the phasing of the ROM such that the location of the amplitude peak in the time domain agrees with SEOBNRv2 to good accuracy.

One can define the notion of \emph{time} from the frequency domain phasing by analogy with \emph{frequency} in the time domain (in a stationary phase sense) as
\begin{equation}
  t_\mathrm{FD}(Mf) / M = \frac{1}{2\pi} \frac{d\phi}{d (Mf)}.
\end{equation}
At low frequency this function is well approximated by a power law~\cite{PhysRevD.50.R7111} 
$t_\mathrm{FD}(Mf) / M \propto (Mf)^{-8/3}$ 
and flattens out as the SEOBNRv2 ringdown frequency $f_\mathrm{rng}$, the highest frequency in the corresponding time domain waveform, is  approached. Thus I take $f_\mathrm{rng}$ as a natural reference point and compute the time elapsed from a starting frequency $f_\mathrm{start}$ until $f_\mathrm{rng}$ which demarcates the end of the waveform.
\begin{equation}
  \Delta T / M = t_\mathrm{FD}(f_\mathrm{rng}) / M - t_\mathrm{FD}(f_\mathrm{start}) / M
\end{equation}

To provide the inverse function, frequency as a function of time, $Mf(t_\mathrm{FD})$ 
I generate data for $t_\mathrm{FD}(Mf)$ on a double-logarithmic grid, spline interpolate the transposed data (frequency as a function of time) to find $I[\log Mf](\log t_\mathrm{FD}/M)$ and obtain 
\begin{equation}
Mf(t_\mathrm{FD}) = \exp(I[\log Mf](\log t_\mathrm{FD}/M))
\end{equation}
In order to avoid pathologies close to merger where $t_\mathrm{FD}$ is not necessarily monotonically decreasing, the double logarithmic grid extends from the ROM starting frequency until half of the ringdown frequency.

In the ROM construction an implicit choice was made for the time and phase of coalescence. These parameters are maximized over when computing the match and thus have not been discussed so far in this paper. GW searches and parameter estimation expect the amplitude peak in the time domain to occur within a prescribed time window and customarily set the origin of time to the location of the peak. To ensure that the inverse Fourier transform of the ROM follows this convention I apply a time shift to the Fourier domain phasing such that $d\phi / d(Mf) \vert_{Mf_\mathrm{rng}} = 0$.

\begin{figure}[htbp]
  \centering
    \includegraphics[width=.49\textwidth]{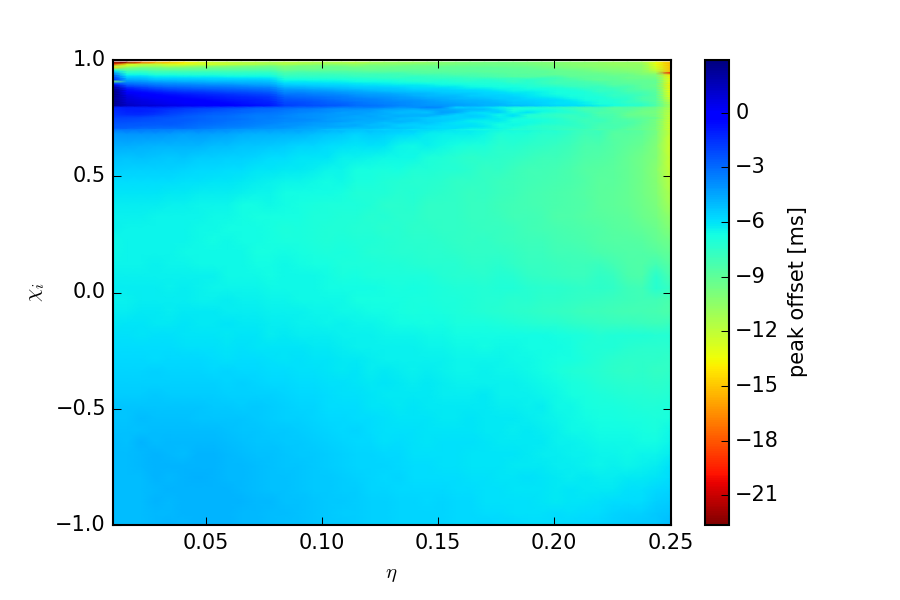}
  \caption{Offset between the time where the amplitude in the time domain peaks for SEOBNRv2 and ROM. If the offset is negative (positive) the peak in the ROM precedes (follows) the peak in SEOBNRv2. The offset is given in milliseconds for configurations at a total mass of $100 M_\odot$ as a function of the symmetric mass-ratio $\eta$ and equal aligned spins $\chi_i$.}
  \label{fig:plots_peak_diff_equal_spin_combined_pcol}
\end{figure}

Fig.~\ref{fig:plots_peak_diff_equal_spin_combined_pcol} shows that the maximum error in the location of the time domain amplitude peak is about $25 \mathrm{ms}$ at a total mass of $100 M_\odot$. The ROM peak occurs up to $\sim 20 \mathrm{ms}$ too early for very high aligned spin systems and close to equal mass systems at moderate to high spin. For moderate spins and mass-ratios the peak offset is about $- 10 \mathrm{ms}$. The maximum peak time error is similar for unequal spin configurations. This deviation scales linearly with the total mass and is within the time windows used by searches and parameter estimation within the LAL codes~\cite{LAL-web} used in the LSC~\cite{LSC_web}.

\section*{Acknowledgements}
I thank Mark Hannam, Andrea Taracchini, Alessandra Buonanno, Frank Ohme, Prayush Kumar, Harald Pfeiffer and Chad Galley for useful discussions and comments.\\

MP was in part supported by Science and Technology Facilities Council grant ST/I001085/1.
SEOBNRv2 waveforms were generated at Advanced Research Computing (ARCCA) at Cardiff.

\section*{References}

\bibliography{ROMv2}

\end{document}